\documentclass[a4paper,10pt]{article}
\usepackage[utf8x]{inputenc}
\usepackage[numbers]{natbib}
\usepackage{amsmath}
\usepackage{amssymb}
\usepackage{dsfont}
\usepackage{color}
\usepackage[dvips]{graphics}
\usepackage[dvips]{graphicx}
\usepackage{psfrag}
\usepackage{multirow}
\usepackage{authblk}

\title{Numerical equilibria with pressure anisotropy and incompressible plasma 
rotation parallel to the magnetic field}
\author{G. Poulipoulis}
\author{ G. N. Throumoulopoulos}
\affil{Section of Astrogeophysics, Physics Department, University of Ioannina, 451 10 Greece}
\date{}

\begin{document}
\bibliographystyle{unsrt}

\maketitle

\begin{abstract}
It is believed that plasma rotation can affect the transitions to the advanced confinement regimes in tokamaks, such as the High 
confinement mode (H-mode) and the formation of Internal Transport Barriers. In addition, in order to achieve fusion temperatures 
modern tokamaks rely on auxiliary heating methods. These methods deposit energy into the charged particles in a specific 
direction and thus generate significant pressure anisotropy in the plasma. For incompressible rotation with pressure anisotropy 
the equilibrium is governed by a Generalized Grad-Shafranov (GGS) equation and a decoupled  Bernoulli-type equation for the 
effective pressure, $\bar{p}=(p_\parallel+p_\perp)/2$, where $p_\parallel$ ($p_\perp$) is the pressure tensor element parallel 
(perpendicular) to the  magnetic field. In the case of plasma rotation parallel to the magnetic field the GGS equation can be 
transformed  to one equation identical in form with the GS equation. In this study by making use of the aforementioned property 
of the GGS equation for parallel plasma rotation we constructed numerical equilibria by extending HELENA, an equilibrium fixed-
boundary solver. The code solves the GGS equation for a variety of the two free-surface-function terms involved for arbitrary 
Alfv\'en Mach number and anisotropy functions. We have constructed ITER-like diverted-boundary equilibria and examined their properties. 
In particular we examined the impact of rotation and anisotropy on certain equilibrium quantities. The main conclusions are that 
the addition of pressure anisotropy to rotation allows the profile shaping of the equilibrium quantities in much more extent thus 
favouring the confinement and allows extension of the parametric space of the Mach number corresponding to higher values. Furthermore, 
the impact of pressure anisotropy in the equilibrium quantities is stronger than that of the rotation, for all quantities, but the 
effective pressure. For the pressure components the impact of the pressure anisotropy is the same regardless of whether the power is deposited 
parallel or perpendicular to the magnetic surfaces, thus implying that there is no preferable heating direction, 
while for the current density, the heating parallel to the magnetic surfaces seems to be beneficial for the current-gradient driven instabilities.
\end{abstract}

\section{Introduction}

In a previous work \cite{2016PhPl...23g2507P} we extended the equilibrium code HELENA \cite{huysmans1991IsobicHerelesolGraequ} to stationary equilibria with rotation parallel to the magnetic field based on experimental and theoretical evidence that plasma flow has impact on the equilibrium, stability and transport 
properties of fusion plasmas. Plasma rotation is associated with the appearance of highly peaked density, pressure and 
temperature profiles, the  suppression of some instabilities and the creation 
of transport barriers either in the H-mode or in discharges with Internal Transport Barriers (see for 
example \cite{gunter2000SimatthigeleiontemdisinttrabarASDupg}, 
\cite{romanelli2011OveJETres} and the review papers 
\cite{itoh1996rolelefiecon}-\cite{challis2004useinttrabartokpla_2}). 
%
 Rotation can be the result of an external source such as electromagnetic power 
and neutral beam injection used for plasma heating and current drive or can manifest itself (intrinsic flows).
It is believed that the reduction of turbulence and the mode decorrelation is the mechanism trough which flow affects the 
confinement. In addition to the flow itself, recent evidence indicates that the spatial variation of the flow affects more 
strongly the confinement, and thus making plasma rotation a significant ingredient for the exploitation of the future big 
machines as ITER and later DEMO in which due to the large plasma volume it would be difficult to induce large flow 
velocities. In these cases it is possible that the intrinsic rotation in these machines can be important 
\cite{solomon2007Momconlowtor}. It is proposed that in JET the driving mechanism for the appearance of intrinsic rotation 
is the pressure gradient \cite{eriksson1997TorrotICRH-mJET}. 

The heating methods used to drive plasma rotation also deposit energy into the 
charged particles in a specific direction and therefore generate significant pressure 
anisotropy in the plasma \cite{2007NucFu..47S.264F, 2001PPCF...43.1441Z, 2011PPCF...53g4021H, 1401.5520v2, 2010PPCF...52f5001P}, 
thus modifying the momentum conservation equation and ultimately affecting the 
equilibrium and stability properties. The magnitude of anisotropy can be 
significant. Depending on the direction of the energy deposition the pressure 
component parallel to that direction increases. For example in a MAST NBI discharge the 
ratio $p_\parallel / p_\perp$ has been found to be 1.7 \citep{2011PPCF...53g4021H} and for a JET 
ICRH discharge the ratio $p_\perp / p_\parallel$ has been found as high as 2.5 
\citep{2001PPCF...43.1441Z}, with $p_\parallel$ and $p_\perp$ the pressure parallel and 
perpendicular to the magnetic field lines.

Equilibrium is the starting point for stability and transport studies. 
For axisymmetric systems, as tokamaks, the governing equation is the so-called Grad-Shafranov (GS) equation whose analytic solutions, as the 
Solov\'ev one, have been found and used for equilibrium and stability studies. 
These analytic solutions are subject to some limitations which only numerical 
solutions can lift. To this end, fixed and free boundary equilibrium codes have 
been developed to solve the equation in realistic situations, i.e. for realistic choices of the boundary (or the currents 
in the coils for free boundary codes) and for the respective free functions, based on information from experimental 
data. Specifically here, we refer to the HELENA code, a fixed boundary solver of the GS equation using finite elements, which is used 
in the present study and further details are given in Sec. 3. In the presence of plasma rotation and pressure anisotropy 
the equilibrium is governed by a generalised  Grad-Shafranov (GGS) equation together with a Bernoulli-type equation 
involving the effective pressure \cite{morozov1980Steplaflomagfie,hameiri1983equstarotpla, 2016PPCF58d5022E,
2009PPCF...51h5011C}. 

From a mathematical point of view, for compressible flows the GGS equation can be either elliptic or hyperbolic. The 
transition depends on certain critical values of the poloidal velocity. It must be pointed out that due 
to axisymmetry of the said configurations, the toroidal velocity is inherently incompressible. In the case of compressible 
flows the GGS equation and the Bernoulli equation are coupled through the density which, in that case, is not a surface 
quantity. In the first elliptic region, which is experimentally accessible\cite{velocity.scaling, mcclements2010steady}, 
and many codes have been developed to solve the system of these two coupled equations as DIVA 
\cite{semenzato1984ComsymideMHDfloequ, strumberger2005NumMHDstastutorrotvisreswalcurhol_2}, FINESSE 
\cite{belien2002FINAxiMHDEquFlo}, FLOW \cite{guazzotto2004Numstutokequarbflo}, including a version of FLOW that 
takes into account non-thermal populations \cite{2009PPCF...51c5014H}. In order to close the system in the 
aforementioned codes an adiabatic or isothermal equation of state is adopted. These equations of state are associated 
with either isentropic or isothermal magnetic surfaces, respectively. The problem of equilibrium with anisotropic pressure and 
toroidal rotation is examined by extension of HELENA \cite{Qu:2014:0741-3335:75007} or EFIT++ 
\cite{Fitzgerald:2013:0029-5515:113040}. For the incompressible flow case the consequence is that the density is uniform 
on the magnetic surfaces, thus the GGS equation (Eq. (\ref{1})) becomes elliptic and decouples from the Bernoulli 
equation. In the case of a fixed boundary, convergence is guaranteed under the requirement of monotonicity for the 
free functions \cite{courant1966Metmatphy}. A code that also assumes density uniform on magnetic surfaces is TRANSP 
\cite{budny1995SimalpparTFTDTsuphigfuspow}. Deviations of density on magnetic surfaces have been observed experimentally, 
thus the use of both compressible in incompressible assumptions in codes contribute in better understanding.

The aim of this work is to develop further the previous work \cite{2016PhPl...23g2507P} where the fixed boundary equilibrium 
code HELENA was extended by including incompressible plasma flow parallel to the magnetic field by adding pressure anisotropy 
and examine the combined effect of rotation and anisotropy on the equilibrium properties. The code is extended taking advantage 
of the fact that the governing GGS, Eq. (\ref{1}),  under a transformation can be put in a form identical with the static and 
pressure isotropic well known GS equation.

In the following section the GGS equation for plasmas with incompressible flow and pressure anisotropy is reviewed. In Sec. 3 the HELENA 
code for parallel flow and pressure anisotropy is presented and the impact of rotation and pressure anisotropy on certain equilibrium 
quantities is examined on specific constructed equilibria. In Section 4 the main conclusions are presented.

\section{Equilibrium equations}
The equations governing a magnetically confined plasma with incompressible flow 
and pressure anisotropy are the following (\cite{1969PlPh...11..211D} 
and ref 6 therein, \cite{2016PPCF58d5022E}):
\begin{align}
\vec{\nabla}\cdot (\rho \vec{\upsilon})=0 \\
 \rho (\vec{\upsilon}\cdot\vec{\nabla} \vec{\upsilon}) + \vec{\nabla} 
\mathds{P} = \vec{J}\times \vec{B} \label{momeq} \\
\vec{\nabla}\times \vec{B}=\mu_0 \vec{J} \\
\vec{\nabla}\cdot \vec{B}=0 \\
\vec{\nabla} \times \vec{E} =0 \\
\vec{E}+\vec{\upsilon}\times \vec{B}=0 \\
\mathds{P}=p_\perp \mathds{I} +\frac{\sigma}{\mu_0}|\vec{B}| 
\end{align}
where $\rho$ is the mass density, $\vec{\upsilon}$ the plasma velocity, 
$\mathds{P}$
the pressure tensor, $\vec{J}$ the current density, $\vec{B}$ the magnetic
field, $\vec{E}$ the electric field, $\mu_0$ the vacuum permeability and the quantity
\begin{equation}
 \sigma = \mu_0 \frac{p_\parallel - p_\perp}{|\vec{B}|^2}
\end{equation}
measures the pressure anisotropy with respect to the parallel ($p_\parallel$) 
and perpendicular ($p_\perp$) to the magnetic surfaces directions.

Assuming an axisymmetric system and defining an effective pressure: 
$$
\overline{p}=\frac{p_\parallel +p_\perp}{2}
$$
one obtains the following Generalized Grad-Shafranov equation \cite{2016PPCF58d5022E} 
\begin{align}
 (1-\sigma - M_p^2) \Delta^\star \psi +
         \frac{1}{2}(\sigma-M_p^2)^\prime |\nabla \psi|^2
                     + \frac{1}{2}\left(\frac{X^2}{1-\sigma 
- M_p^2}\right)^\prime 
\nonumber  \\
+\mu_0 R^2 \overline{p}_s^\prime + \mu_0 \frac{R^4}{2}\left[ 
\frac{\rho(\Phi^\prime)^2}{1-\sigma - M_p^2}\right]^\prime
    = 0
 \label{1}
 \end{align}
 Here, $\psi(R,z)$ is the poloidal magnetic flux function associated to the magnetic 
surfaces, where ($R,\phi, z$) are cylindrical coordinates; $\phi$ is the ignorable coordinate; the function $M_p(\psi)$ is
 the Alfv\'en Mach number of the fluid velocity along the poloidal direction; $X(\psi)$ is a surface quantity that refers 
 to the toroidal magnetic field, $B_\phi=I/R$. The relation that connects these two quantities is
$I=(X-R^2\sqrt(\rho)M_p\Phi')/(1-\sigma - M_p^2)$; with $\Phi(\psi)$ being the electrostatic potential. In the static case
$\overline{p}_s(\psi)$ coincides with the effective pressure; $B$ is the magnetic field modulus depending on surface 
quantities and the radial coordinate; $\Delta^\star=R^2\nabla\cdot(\nabla/R^2)$; while derivatives with respect to $\psi$ are 
denoted by the prime.
As mentioned before, a consequence of incompressibility is that the density $\rho(\psi)$ becomes a surface quantity leading to the
decoupling of the Bernoulli equation from the GGS (\ref{1}):
\begin{equation}
 \overline{p}=\overline{p}_s(\psi) - \varrho \left( \frac{\upsilon^2}{2} - 
\frac{R^2 (\Phi^\prime)^2}{1-M_p^2}\right)
                          \label{2}
 \end{equation}
 with $\upsilon$ being the velocity modulus. The functions
$M_p(\psi)$, $\sigma(\psi)$, $X(\psi)$, $p_s(\psi)$, $\rho(\psi)$ and 
$\Phi(\psi)$ are free. In addition, the parallel and perpendicular components of the pressure tensor are given by:
\begin{align}
 p_\perp = \overline{p} - \sigma \frac{B^2}{2\mu_0} \label{pv} \\
 p_\parallel = \overline{p} + \sigma \frac{B^2}{2\mu_0} \label{ppar}
 \end{align}
Details for the derivation of Eq. (\ref{1}) are given in 
\cite{2016PPCF58d5022E}. The main steps are first to express the divergence free fields ($\vec{B}$, $\vec{J}$ and $\rho 
\vec{v}$) in terms of scalar quantities and second, project the momentum equation (\ref{momeq}),
and Ohm's law, along the toroidal direction, $\vec{B}$ and $\vec{\nabla}\psi$. 
The projections yield four first integrals in the form of surface quantities
and Eqs. (\ref{1}) and \ref{2}.

Applying the transformation \cite{1993NucFu..33..963C} 
\begin{equation}
u(\psi) = \int_{0}^{\psi}\left\lbrack 1 - \sigma(f) -
M_p^{2}(f)\right\rbrack^{1/2} df
\label{3}
\end{equation}
to Eq. (\ref{1}) it becomes
\begin{align}
  \Delta^\star u
+ \frac{1}{2}\frac{d}{du}\left(\frac{X^2}{1- \sigma - M_p^2}\right)  +
\mu_0 R^2\frac{d \overline{p}_s}{d u} \nonumber \\
+ \mu_0\frac{R^4}{2}\frac{d}{du}\left[(1- \sigma) \rho\left(\frac{d 
\Phi}{du}\right)^2\right] = 0
 \label{4}
\end{align}
The latter equation does not contain a quadratic term as $|{\bf\nabla}u|^{2}$. Once a solution of (\ref{4}) 
is obtained, the equilibrium can be completely  constructed with calculations 
in the $u$-space by employing (\ref{3}), and  the inverse transformation
\begin{equation}
\psi(u) = \int_{0 }^{u}\left\lbrack 1 - \sigma(f) -
M_p^{2}(f)\right\rbrack^{-1/2} df
                                            \label{5}
\end{equation}
Specifically, the correspondence between  $u$-space and the $\psi$-space for some quantities are:  
\begin{align}
\overline{p}=\overline{p}_s(u)-\varrho(u)\left[\frac{\upsilon^2}{2}
-\frac{R^2(1-\sigma)}{1-\sigma -M_p^2} \left(\frac { d\Phi(u) } { du }
\right)^2\right] \label{pres1} \\
\vec{B}=I(\psi)\vec{\nabla}\phi -\vec{\nabla}\phi\times\vec{\nabla}\psi= 
I(u)\vec{\nabla}\phi-\frac{d\psi}{du}\vec{\nabla}\phi\times\vec{\nabla}u 
\label{magfiel} \\
\vec{J}=\frac{1}{\mu_0}\left(-\Delta^*\psi\vec{\nabla}\phi+\vec{
\nabla}\phi\times \vec{\nabla}I(\psi)\right)= \nonumber \\
\frac{d\psi}{du}R^2\vec{\nabla}\left(\frac{\vec{\nabla}u}{R^2}\right)+ 
\vec{\nabla} u \cdot \vec{\nabla}\frac{d\psi}{du} +\frac{dI(u)}{du}\vec{ 
\nabla}\phi\times \vec{\nabla}u \label{curden}\\
\vec{E}=-\vec{\nabla} \Phi= -\frac{d \Phi(\psi)}{d \psi} \vec{\nabla} \psi= 
-\frac{d \Phi(u)}{d u} \vec{\nabla} u
\end{align}
For flows aligned to the magnetic field, ($\Phi^\prime=0$), Eq. (\ref{4}) takes the form of the 
usual GS equation and can be shown that the poloidal, toroidal and total velocity Alfv\'{e}n Mach numbers 
are exactly equal; thus we drop the subscript in the Mach number. One additional surface quantity, $K$, can be obtained by setting
\begin{equation}
\varrho\vec{\upsilon}=K\vec{B}, \label{veloc}
\end{equation}
Applying the divergence operator and taking into account the continuity equation, 
$\vec{\nabla}\cdot\left(\varrho\vec{\upsilon}\right)=0$, one obtains 
$\vec{\nabla}K\cdot\vec{B}=0 $ as was shown in \cite{throumoulopoulos2003axiresmagequflofrePfidif}.
Finally, the Bernoulli Eq. (\ref{2}) with the aid of Eq. (\ref{veloc}), becomes
\begin{equation}
 \overline{p}=\overline{p}_s(\psi) - \frac{1}{2\mu_0}M^2B^2(\psi, R)= \overline{p}_s(u) 
- \frac{1}{2\mu_0}M^2B^2(u, R)  \label{pres2}
\end{equation}

\section{Numerical equilibria with parallel plasma rotation and pressure anisotropy}

In order to examine the impact of parallel plasma rotation in combination with the pressure anisotropy we constructed 
numerical equilibria by remapping and making appropriate use of the code HELENA. The code is a fixed boundary equilibrium 
solver which solves the static GS equation written as:
\begin{equation}
\Delta^*\psi = -F\frac{dF}{d\psi} - \mu_0R^2\frac{dP}{d\psi} = -\mu_0Rj_{tor}
\label{eq5}
\end{equation}
The code makes use of isoparametric bi-cubic Hermite finite elements to solve the above equation by 
employing the Galerkin method, which is a non-linear iteration scheme, and using straight-field-line coordinates.
The boundary condition consists of specific values for the function $\psi$ on a predefined curve which for the code coincides 
with the last closed flux surface of its computational domain. The technique has proved to produce high quality results with 
fast convergence \cite{KonzHELENA}.

The following mapping can be established by comparison of Eq. (\ref{4}) for parallel plasma rotation ($\Phi^\prime=0$)  with 
Eq. (\ref{eq5}): 
\begin{align}
\psi \longleftrightarrow u 
\label{eq6} \\
F\frac{dF}{d\psi} \longleftrightarrow \frac{1}{2}\frac{d}{du}\left( 
\frac{X^2}{1 -\sigma - M^2}\right) \label{eq7} \\
P(\psi) \longleftrightarrow \overline{p}_s(u) 
\label{eq8}
\end{align}
On the basis of this correspondence, it is expected that the use of HELENA for the computation of stationary equilibria for parallel 
plasma flow and pressure anisotropy is possible, with the following clarification. The solver of the code remains unchanged, 
but the input/output quantities to it no longer refer to the $\psi$-space. A correspondence between the $\psi$-space and the 
$u$-space of the input and the calculated by the solver quantities respectively is required. For the minimum set of the basic 
quantities, the aforementioned mapping is:
\begin{align}
P_{\mbox{\scriptsize HELENA}}\longleftrightarrow \overline{p}_s \label{eq8a} \\
F_{\mbox{\scriptsize HELENA}}\longleftrightarrow \frac{X}{\sqrt{1- \sigma - 
M^2}} \\
\psi_{\mbox{\scriptsize HELENA}}\longleftrightarrow u \label{eq8c}
\end{align}
The mapping for the magnetic field, the current density and the pressure by making use of Eqs. (\ref{eq8a})-(\ref{eq8c}) on 
(\ref{magfiel}), (\ref{curden}) for $\Phi^\prime=0$ and (\ref{pres2}) is:
\begin{align}
\addtolength{\itemsep}{-5mm}
\vec{B}=\frac{F_{\mbox{\scriptsize
HELENA}}}{\sqrt{1- \sigma - M^2}}\vec{\nabla}\phi-\frac{1}{\sqrt{1- \sigma - 
M^2}}
\vec{\nabla}\phi\times\vec{\nabla}u \label{eq:9} \\
\vec{J}=\left[\frac{-1}{\sqrt{1- \sigma - 
M^2}}\Delta^*u+\frac{1}{2}\frac{1}{(1- \sigma - M^2)^{3/2}
} \frac{d(\sigma + M^2)}{du} |\vec{\nabla}u|^2\right]\vec{\nabla}\phi+  \nonumber 
\\
\frac{d}{du}\left(\frac{F_{\mbox{
\scriptsize HELENA}}}{
\sqrt{1- \sigma  - M^2}}\right) \vec{\nabla} \phi\times\vec{\nabla}u 
\label{eq:10} \\
 \overline{p}=P_{\mbox{\scriptsize HELENA}}-\frac{1}{2 \mu_0
R^2}\frac{M^2}{1-M^2}\left(F_{\mbox{\scriptsize HELENA}}^2+|\vec{\nabla}
u|^2\right) 
\label{eq:11}
\end{align}
where the subscript HELENA refers to the quantities computed by the solver of the Grad-Shafranov equation.

A point of interest is that owing to the fact that the transformation (\ref{3}) just relabels the magnetic surfaces and 
that HELENA is a fixed boundary solver- therefore the ``radial" dependence of the magnetic field is not affected by the 
plasma rotation and pressure anisotropy- the safety factor $q$ is flow and anisotropy independent, as long as the input 
to the solver remains fixed. One must clarify that the input to the solver is not the same as the input to the code, since 
the latter depends on the integral transformation, in the presence of rotation and pressure anisotropy. Specifically, the 
safety factor is given by the relation 
\cite{ideal_magnetohydrodynamics}
\begin{equation}
 q(\psi)=\frac{1}{2\pi}\int^{2\pi}_0 \left(\frac{r B_\phi}{R 
B_\theta}\right)_S d\theta=\frac{F(\psi)}{2\pi}\oint\frac{ d\ell_p}{ R^2 
B_p}=\frac{F(\psi)}{2\pi|\nabla\psi|}\oint\frac{ d\ell_p}{ R}
 \label{eq:q1}
\end{equation}
The integration in the last two expressions is performed along the curve of a magnetic surface in the poloidal plane and $r$, 
$\theta$ are cylindrical coordinates of a system with its origin located at the position of 
the magnetic axis, with $d\ell_p=\sqrt{(dr)^2+(rd\theta)^2}$, 
$B_p=\sqrt{B_r^2+B_\theta^2}=|\nabla\psi|/R$. Applying the integral transformation on $F(\psi)$ and $\nabla\psi$ and 
taking into account (\ref{eq:9}) for the components of the magnetic field we get
\begin{equation}
 q(\psi)=q(u)= \frac{F(u)}{2\pi|\nabla u|}\oint\frac{ d\ell_p}{ R}
 \label{eq:q2}
\end{equation}
Therefore, as long as the input to the solver remains the same, so does the safety factor. 

\begin{figure}[ht!]
\begin{center}
\psfrag{ppr}{$\overline{p}'$}
\psfrag{psi}{$\psi_{norm}$}
\includegraphics[scale=0.46]{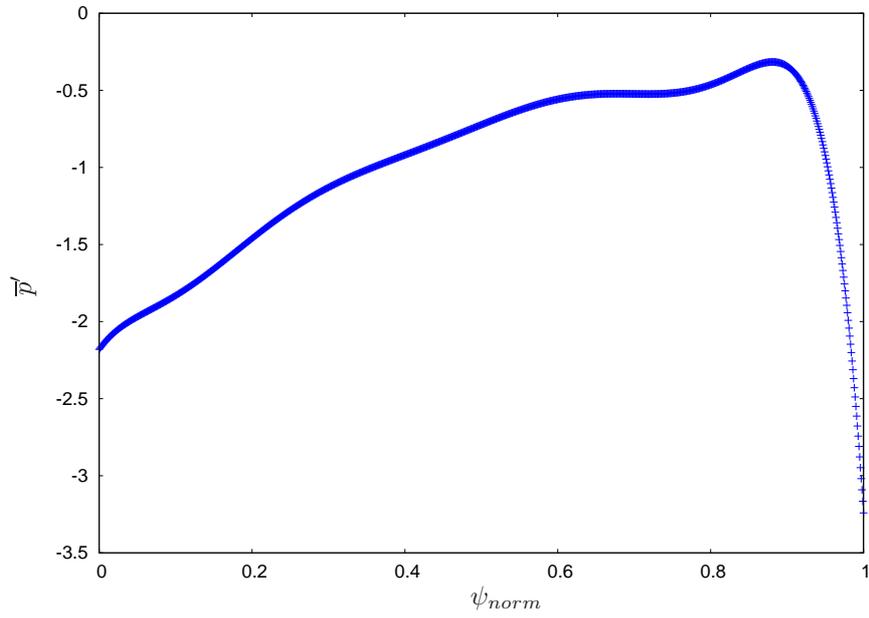}
\caption{The input profile of the derivative of the effective pressure with respect to a normalized $\psi$ defined as $\psi/\psi_{boundary}$.}
\label{fig:ipp}
\end{center}
\end{figure}

\begin{figure}[ht!]
\begin{center}
\psfrag{ffpr}{$FF'$}
\psfrag{psi}{$\psi_{norm}$}
\includegraphics[scale=0.46]{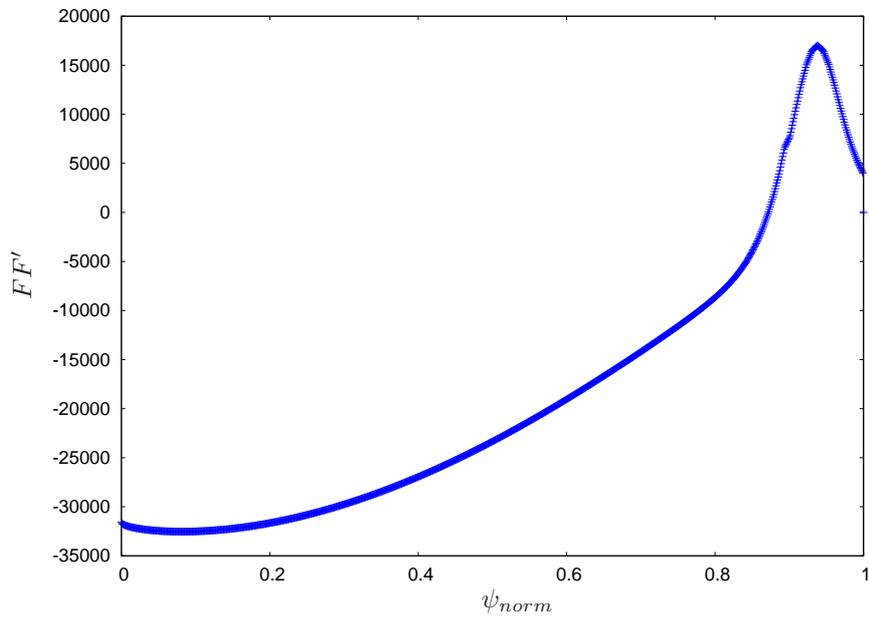}
\caption{The input profile of $FF'$ used in the runs of the code with respect to normalized $\psi$.}
\label{fig:ffp}
\end{center}
\end{figure}

\begin{figure}[ht!]
\begin{center}
\psfrag{R}{$R(m)$}
\psfrag{z}{$z(m)$}
\includegraphics[scale=0.28]{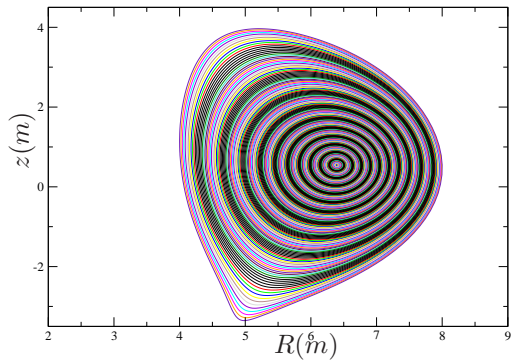}
\caption{The magnetic surfaces of a diamagnetic  equilibrium associated with 
the values of Table \ref{tab:0} and the input profiles of Figs. \ref{fig:ipp} 
and and \ref{fig:ffp} . 
The magnetic axis is located at ($R_{a}=6.394$ m, 
$z_{a}=0.54585$ m)  where the toroidal magnetic field is 4.969 T with respective 
vacuum value 5.3 T.}
\label{fig:surf}
\end{center}
\end{figure}

It is worth pointing out that the solutions of (\ref{eq5}), and therefore of the 
extended code, hold for arbitrary functions of the Mach number $M(u)$, anisotropy $\sigma(u)$ and density $\varrho(u)$. 
By varying the input quantities of the modified code we obtained a number of equilibria. As an example, the magnetic surfaces 
of an ITER-like configuration with input values summarized in Table \ref{tab:0} and input functions for the quantities $P$ 
and $FF'$ those in Figs. \ref{fig:ipp} and \ref{fig:ffp}, respectively, are presented in Fig. \ref{fig:surf}.
\begin{table}
\centering
\caption{Input values of the basic quantities used for the equilibria 
solutions.}
\begin{tabular}{| c | c | c | c |}
\hline 
$R_0$ & $B_{\phi 0}$ & $I_{plasma}$ & $\psi_{axis}$ \\ \hline
6.00 m  & 5.3 T & 15.1 MA &  0.0 Wb \\
 \hline
 \end{tabular}
 \label{tab:0}
\end{table}

In order to calculate all the equilibrium quantities we modelled the free functions $M^2(u)$ and $\sigma(u)$, 
in two different ways each e.g., 
\begin{eqnarray}
M^2& =& M_0^2 \left(u^m - u_b^m\right)^n
\label{prof1} \\
M^2&=& C \left[\left(\frac{u}{ u_b}\right)^m\left( 
1-\left(\frac{u}{ u_b}\right)\right)\right]^n
\label{prof2} \\
\sigma& =& \sigma_0 \left(u^k - u_b^k\right)^\ell \label{prof3} \\
 \sigma&=& D \left[\left(\frac{u}{ u_b}\right)^k\left( 
1-\left(\frac{u}{ u_b}\right)\right)\right]^\ell
\label{prof4} 
\end{eqnarray}
 where
$$
C=M_0^2\left(\frac{m+n}{m}\right)^m \left(\frac{n}{m+n}\right)^n
$$
and
$$
D=\sigma_0\left(\frac{k+\ell}{k}\right)^k \left(\frac{\ell}{k+\ell}\right)^\ell
$$
In this notation, $ u_b$ refers to the plasma boundary;  the free parameters
$M_0^2$ and $\sigma_0$ correspond to the maximum value of $M^2$ and $\sigma$; 
and $m$, $n$ are related to flow shear and the position of the 
maximum $M^2$; $k$ and $\ell$ are related to the spatial anisotropy variation (shear)
and the position of $\sigma_0$. In particular, (\ref{prof1}) and (\ref{prof3}) are  peaked on- while (\ref{prof2}) and 
(\ref{prof4}) peaked off-axis. This specific choice is associated with respective auxiliary heating of tokamaks. For parallel 
rotation the density does not appear explicitly in the equilibrium equations and hence there is no need to specify it.
The scaling
$
M^2=\frac{v^2}{B^2/(\mu_0 \rho)} \sim \alpha \frac{v_s^2}{B^2/(\mu_0 \rho)}\sim 
\alpha \frac{\gamma P}{B^2/(\mu_0 \rho)}\sim \alpha \beta 
$
where $v_s=\left(\gamma P/\rho\right)^{1/2}$ is the sound speed and 
$\beta=2P/(B^2/\mu_0)$ can be used to estimate $M$. Since the maximum experimental value of $v$ in tokamaks 
is of the order of $v_s$ ($\alpha \sim 0.01-0.1$) and $\beta \sim 0.01$,  the 
experimental values of $M$ lie in the interval ($10^{-2}, 10^{-1}$). In small tokamaks where the torque input can 
produce large plasma flow due to the small volume, the values can be even larger. 
The choice for peaked on- and off-axis of the Mach number function was motivated by experimental 
observations of equilibria with plasma rotation \citep{crombe2005poloidal, 
2001NucFu..41..865S, fiore2012production}, while the numerical values of $M_0$ 
\citep{devries2008ScarotmomconJETpla}, in some cases well above the 
experimental values, are used for illustrative reasons. Similarly, the choice for the respective anisotropy profiles 
is based on physical considerations related to the position where the energy of the heating beams is deposited. In other 
studies (\citep{2010PPCF...52f5001P, 2011PPCF...53g4021H, 1401.5520v2}) the anisotropy was located in the plasma core region. 
In some cases it is possible to have anisotropy peaked off-axis when the heating is focused in some region away from 
the magnetic axis. Regarding the values of $\sigma_0$, it is reminded that in JET has been reported anisotropy as high as 
$p_\perp / p_\parallel \approx 2.5$ \citep{2001PPCF...43.1441Z}.

\begin{figure}[ht!]
\begin{center}
\psfrag{m}{$M$}
\psfrag{r}{$\psi$ (Wb)}
\psfrag{m20223s30252 }{\hspace{0.8cm}\small{Case 1}}
\psfrag{m30452s20223 }{\hspace{0.8cm}\small{Case 2}}
\includegraphics[scale=0.46]{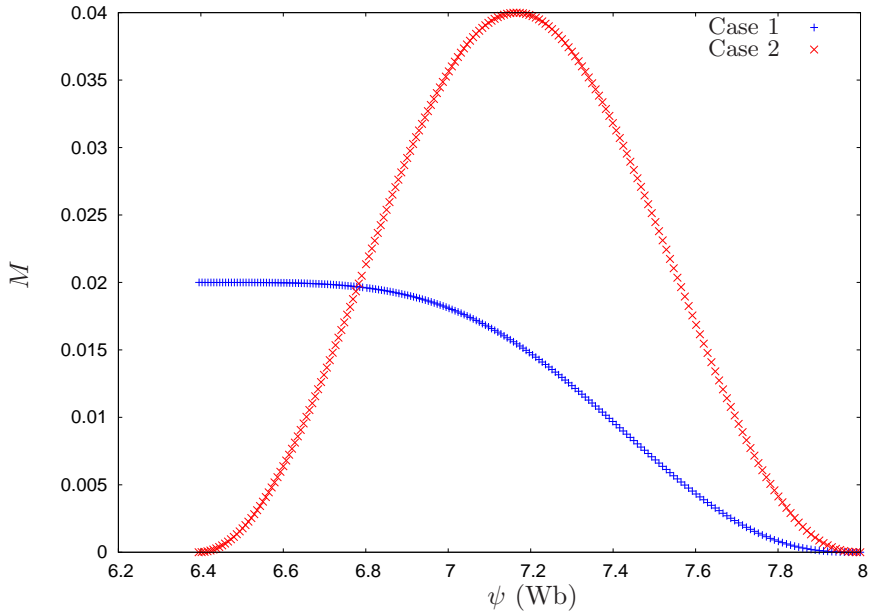}
\caption{Plots of the Mach number profile with respect to $\psi$ used in the calculated 
equilibria. Case 1: (blue \textcolor{blue}{$+$}) peaked on-axis (Eq. (\ref{prof1})) Mach number with $M_0=0.02$, $m=2$, $n=3$; 
Case 2: (red \textcolor{red}{$\times$}) peaked off-axis (Eq. (\ref{prof1})) Mach number with $M_0=0.04$, $m=5$, $n=2$.}
\label{fig:mach1}
\end{center}
\end{figure}

\begin{figure}[ht!]
\begin{center}
\psfrag{s}{$\sigma$}
\psfrag{r}{$\psi$ (Wb)}
\psfrag{m0s20223 }{\hspace{0.35cm}\small{Case 1}}
\psfrag{m0s20226 }{\hspace{0.35cm}\small{Case 2}}
\psfrag{m0s203523 }{\hspace{0.5cm}\small{Case 3}}
\includegraphics[scale=0.46]{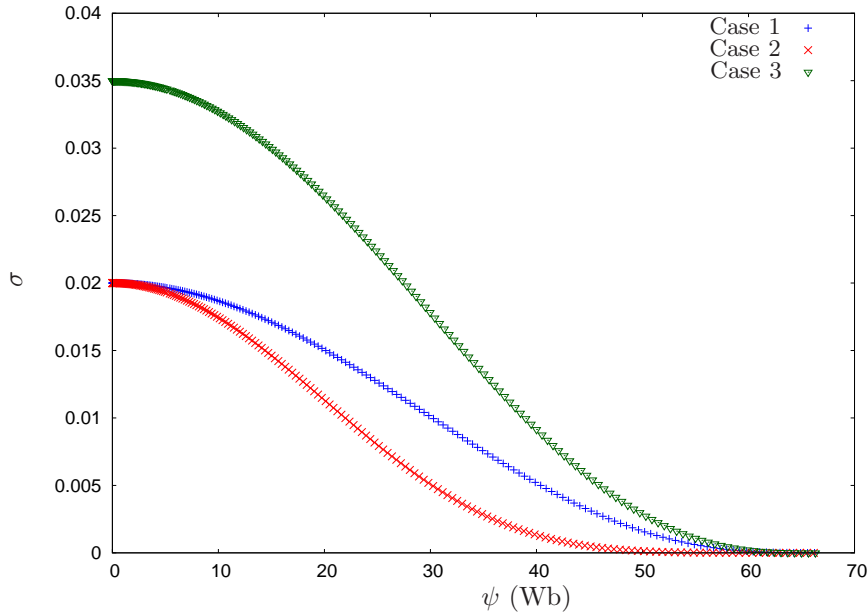}
\caption{Plots of the on-axis peaked $\sigma$ profile (Eq. (\ref{prof3})) with respect to $\psi$  for three cases. Case 1: 
(blue \textcolor{blue}{$+$}) $\sigma_0=0.02$, $k=2$, $\ell=3$; Case 2: (red \textcolor{red}{$\times$}) $\sigma_0=0.02$, 
$k=2$, $\ell=6$; Case 3: (green \textcolor{green}{$\triangledown$}) $\sigma_0=0.035$, $k=2$, $\ell=3$.}
\label{fig:sigma1}
\end{center}
\end{figure}

\begin{figure}[ht!]
\begin{center}
\psfrag{s}{$\sigma$}
\psfrag{r}{$\psi$ (Wb)}
\psfrag{m0s30252 }{\hspace{0.2cm}\small{Case 1}}
\psfrag{m0s300552 }{\hspace{0.35cm}\small{Case 2}}
\psfrag{m0s300524 }{\hspace{0.35cm}\small{Case 3}}
\psfrag{m30424s30241 }{\hspace{0.75cm}\small{Case 4}}
\includegraphics[scale=0.46]{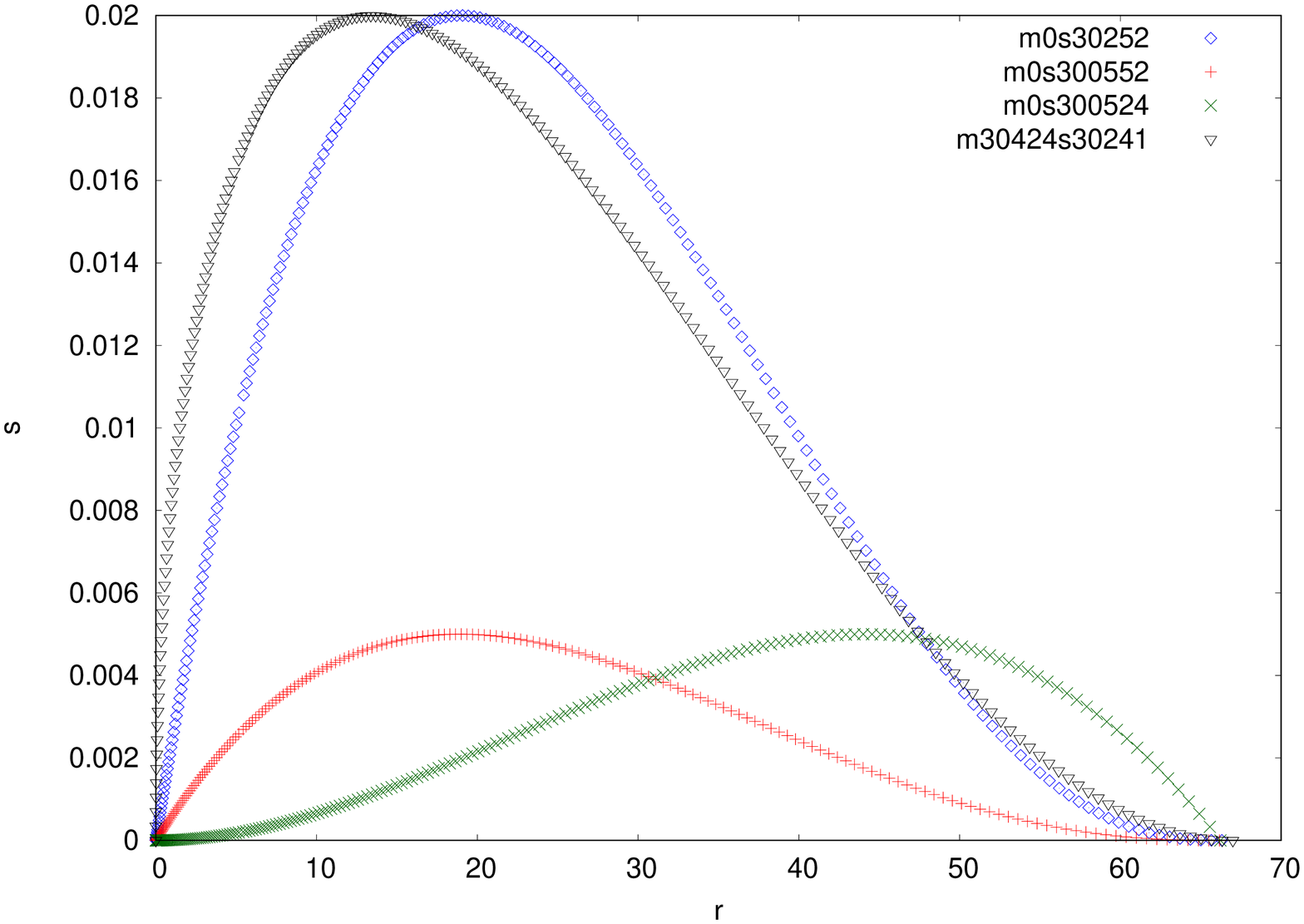}
\caption{Plots of the off-axis peaked $\sigma$ profile (Eq. (\ref{prof4})) with respect to $\psi$ for various cases. Case 1: 
(blue \textcolor{blue}{$+$}) $\sigma_0=0.02$, $k=5$, $\ell=2$; Case 2: (red \textcolor{red}{$\times$}) $\sigma_0=0.005$, $k=5$, 
$\ell=2$; Case 3: (green \textcolor{green}{$\triangledown$}) $\sigma_0=0.005$, $k=2$, $\ell=4$; Case 4: (black \textcolor{black}{$\diamondsuit$}) 
$\sigma_0=0.02$, $k=4$, $\ell=1$.}
\label{fig:sigma2}
\end{center}
\end{figure}

The experimental profiles that are characterized either by a maximum at the magnetic axis or one at a point within the plasma volume 
was the motivation for the specific choice. At the same time the rotation or the pressure anisotropy are localized in a finite 
region of the poloidal plane. However, it should be clarified that the specific choices made for the input profiles do not 
reproduce precisely experimental profiles throughout the poloidal cross-section. Profile examples for the choices (\ref{prof1}) 
and (\ref{prof2}) for $M$ (\ref{prof3}) and (\ref{prof4}) for $\sigma$ by varying the free parameters are given in Figs. 
\ref{fig:mach1}, \ref{fig:sigma1} and \ref{fig:sigma2}, respectively.

By inspection of Eq. (\ref{4}) one expects that the rotation has a weak contribution. However, as already mentioned in Sec.
 1, it seems that the velocity shear is more important than the velocity amplitude for the transition to improved confinement 
modes in tokamaks, a result that was reported in \citep{2016PhPl...23g2507P}, in association with equilibrium profiles compatible with ones
present in configurations with transport barriers. The impact of pressure anisotropy is expected to be 
qualitatively the same as the one of the rotation for $\sigma>0$, though quantitatively it will be stronger than that of the 
rotation due to the fact that the Mach number enters squared in the equations while $\sigma$ enters linearly. In addition, the 
impact of $\sigma$ on equilibrium differs from that of $M$ because, depending on the energy deposition direction, it can take negative 
values. The presence of pressure anisotropy allows larger, though out of the experimental limits for the large 
tokamaks, values for the Mach number. One more point worth noting is that the two free 
quantities ($\sigma$, $M$) can be potentially used for the shaping of the equilibrium profiles, thus 
affecting the stability properties of the configuration, especially for gradient-driven 
instabilities. 

%
%
We examined the effect of plasma flow and pressure anisotropy in some equilibrium quantities by varying the parameters of the 
Mach number profile ($M_0$, $n$, $m$) and the pressure anisotropy function ($\sigma_0$, $k$, $\ell$). We note here that any 
external momentum or energy sources have not been included in the equilibrium equations and therefore the total energy of 
the system is conserved. Sustaining the configuration after having achieved a desired performance and consequently removing 
the external energy and momentum sources is desirable for the operation of a tokamak reactor.

A few general remarks regarding the impact of rotation and pressure anisotropy are that the combination of 
these two quantities allow to shape the equilibrium profiles with great flexibility; specifically, 
the fact that $\sigma$ can become negative broadens the region of permissible values for $M$, thus allowing to access higher 
plasma rotation velocities. This, in turn, leads to stronger impact of the rotation on those equilibrium quantities where, in 
addition to the term $(1 - \sigma - M^2)$, there exist other explicit terms of $M$, $M^2$ or $(M^2)'$ (cf. Eqs. \eqref{4} 
and \eqref{eq:9}-\eqref{eq:11}).  The impact of plasma rotation and pressure anisotropy with respect to the maximum values 
as well as to the shear of the profiles of these quantities is examined in detail as it follows. By varying  
$M_0$ (or $\sigma_0$) on the one hand and $n$, $m$ ($\ell$, $k$) on the other we examine the impact of plasma rotation 
(pressure anisotropy) and its shear on the equilibrium. We will focus mainly on the effect of pressure 
anisotropy and compare it to the one by parallel plasma rotation, while the effect of the latter was individually examined 
in \cite{2016PhPl...23g2507P}.

\begin{figure}[ht!]
\begin{center}
\psfrag{p}{$\overline{p}$ (Pa)}
\psfrag{r}{R(m)}
\psfrag{staticiso }{\small{Case 1}}
\psfrag{m30424s30252 }{\hspace{0.8cm}\small{Case 2}}
\psfrag{m30424s3m0252 }{\hspace{1cm}\small{Case 3}}
\psfrag{m30424s30241 }{\hspace{0.8cm}\small{Case 4}}
\includegraphics[scale=0.46]{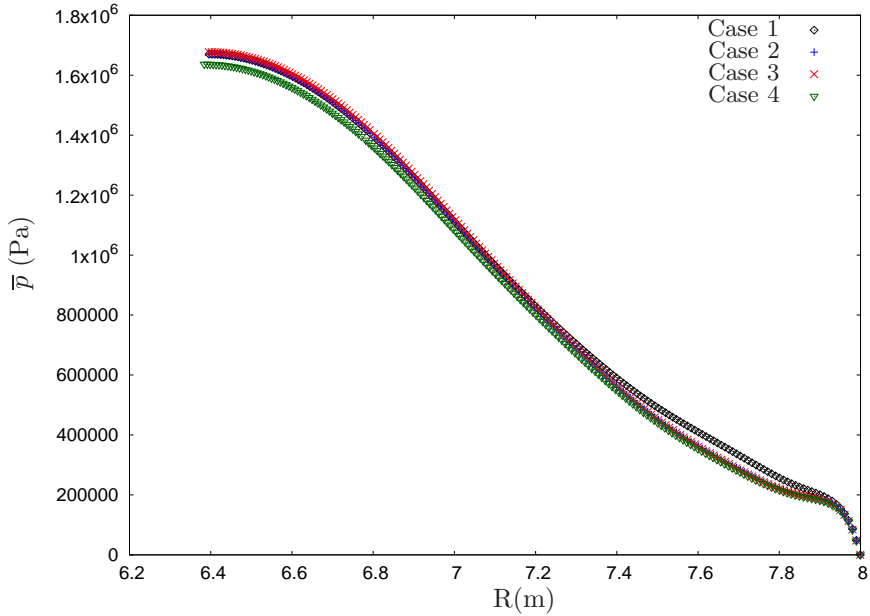}
\caption{Radial profiles of the effective pressure on the mid-plane $z=0$ for various cases of pressure anisotropy and plasma 
rotation. Case 2: (blue \textcolor{blue}{$+$}) peaked off-axis pressure anisotropy with $\sigma_0=0.02$, $k=5$, $\ell=2$; Case 3: 
(red \textcolor{red}{$\times$})  peaked on-axis pressure anisotropy with $\sigma_0=-0.02$, $k=5$, $\ell=2$; Case 4: (green 
\textcolor{green}{$\triangledown$})  peaked on-axis pressure anisotropy with $\sigma_0=0.02$, $k=4$, $\ell=1$. For reference the 
static effective pressure profile (black $\diamondsuit$, Case 1) is also given.}
\label{fig:pres1}
\end{center}
\end{figure}

\begin{figure}[ht!]
\begin{center}
\psfrag{p}{$\overline{p}$ (Pa)}
\psfrag{r}{R(m)}
\psfrag{staticiso }{\small{Case 1}}
\psfrag{m20223s20223 }{\hspace{0.8cm}\small{Case 2}}
\psfrag{m20223s20226 }{\hspace{0.8cm}\small{Case 3}}
\psfrag{m20223s203523 }{\hspace{0.92cm}\small{Case 4}}
\includegraphics[scale=0.46]{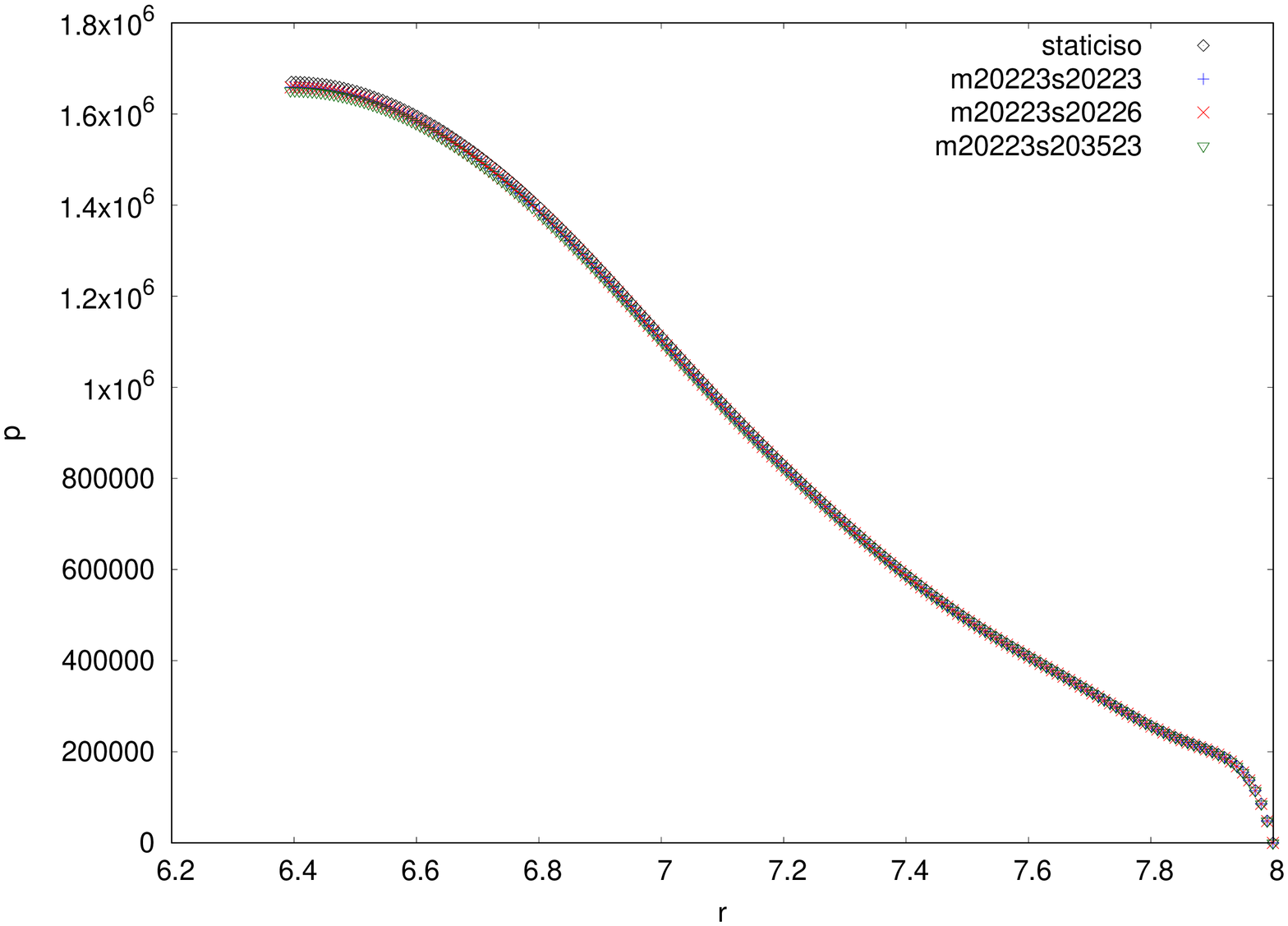}
\caption{Radial profiles of the effective pressure on the mid-plane $z=0$ for various cases of pressure anisotropy and plasma 
rotation. Case 2: (blue \textcolor{blue}{$+$}) peaked on-axis pressure anisotropy with $\sigma_0=0.02$, $k=2$, $\ell=3$; 
Case 3: (red \textcolor{red}{$\times$})  peaked on-axis pressure anisotropy with $\sigma_0=0.02$, $k=2$, $\ell=6$; Case 4: 
(green \textcolor{green}{$\triangledown$})  peaked on-axis pressure anisotropy with $\sigma_0=0.035$, $k=2$, $\ell=3$. In all 
cases, the rotation is peaked on-axis with $M_0=0.02$, $m=2$ and $n=3$. For reference the static effective pressure profile 
(black $\diamondsuit$, Case 1) is also given.}
\label{fig:pres8}
\end{center}
\end{figure}

\begin{figure}[ht!]
\begin{center}
\psfrag{p}{$p_\parallel$ (Pa)}
\psfrag{r}{R(m)}
\psfrag{staticiso }{\small{Case 1}}
\psfrag{m0s20223 }{\hspace{0.25cm}\small{Case 2}}
\psfrag{m0s20226 }{\hspace{0.25cm}\small{Case 3}}
\psfrag{m0s203523 }{\hspace{0.4cm}\small{Case 4}}
\includegraphics[scale=0.46]{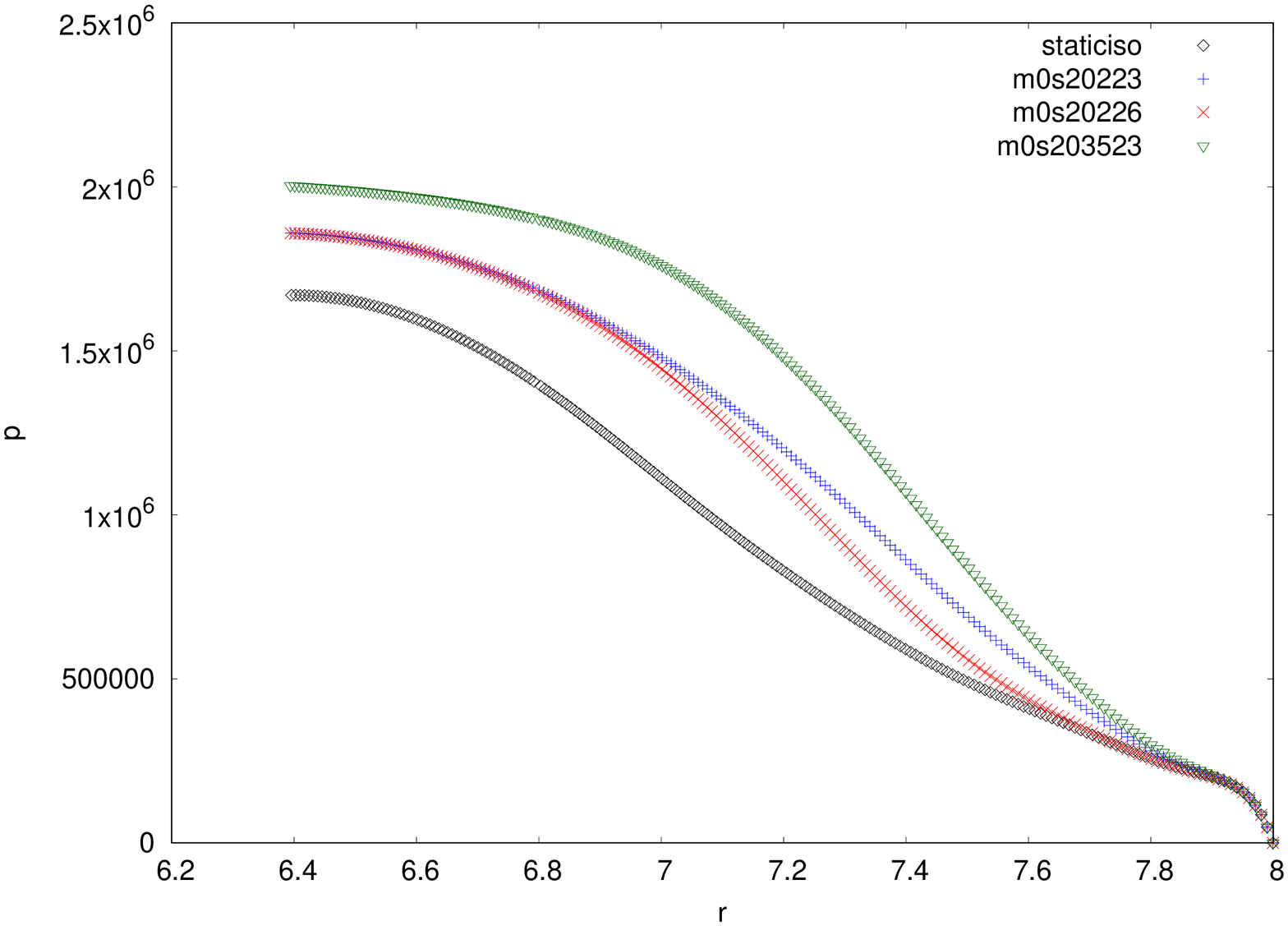}
\caption{Radial profiles of the parallel pressure on the mid-plane $z=0$ for various cases of the pressure anisotropy 
parameters and no plasma rotation.  Case 2: (blue \textcolor{blue}{$+$}) peaked on-axis pressure anisotropy with 
$\sigma_0=0.02$, $k=2$, $\ell=3$; Case 3: (red \textcolor{red}{$\times$})  peaked on-axis pressure anisotropy with 
$\sigma_0=0.02$, $k=2$, $\ell=6$; Case 4: (green \textcolor{green}{$\triangledown$})  peaked on-axis pressure anisotropy 
with $\sigma_0=0.035$, $k=2$, $\ell=3$. The static effective pressure (black $\diamondsuit$, Case 1) is plotted for reference.}
\label{fig:pres2}
\end{center}
\end{figure}

\begin{figure}[ht!]
\begin{center}
\psfrag{p}{$p_\parallel$ (Pa)}
\psfrag{r}{R(m)}
\psfrag{staticiso }{\small{Case 1}}
\psfrag{m20223s30252 }{\hspace{0.8cm}\small{Case 2}}
\psfrag{m20223s3m0252 }{\hspace{1cm}\small{Case 3}}
\includegraphics[scale=0.46]{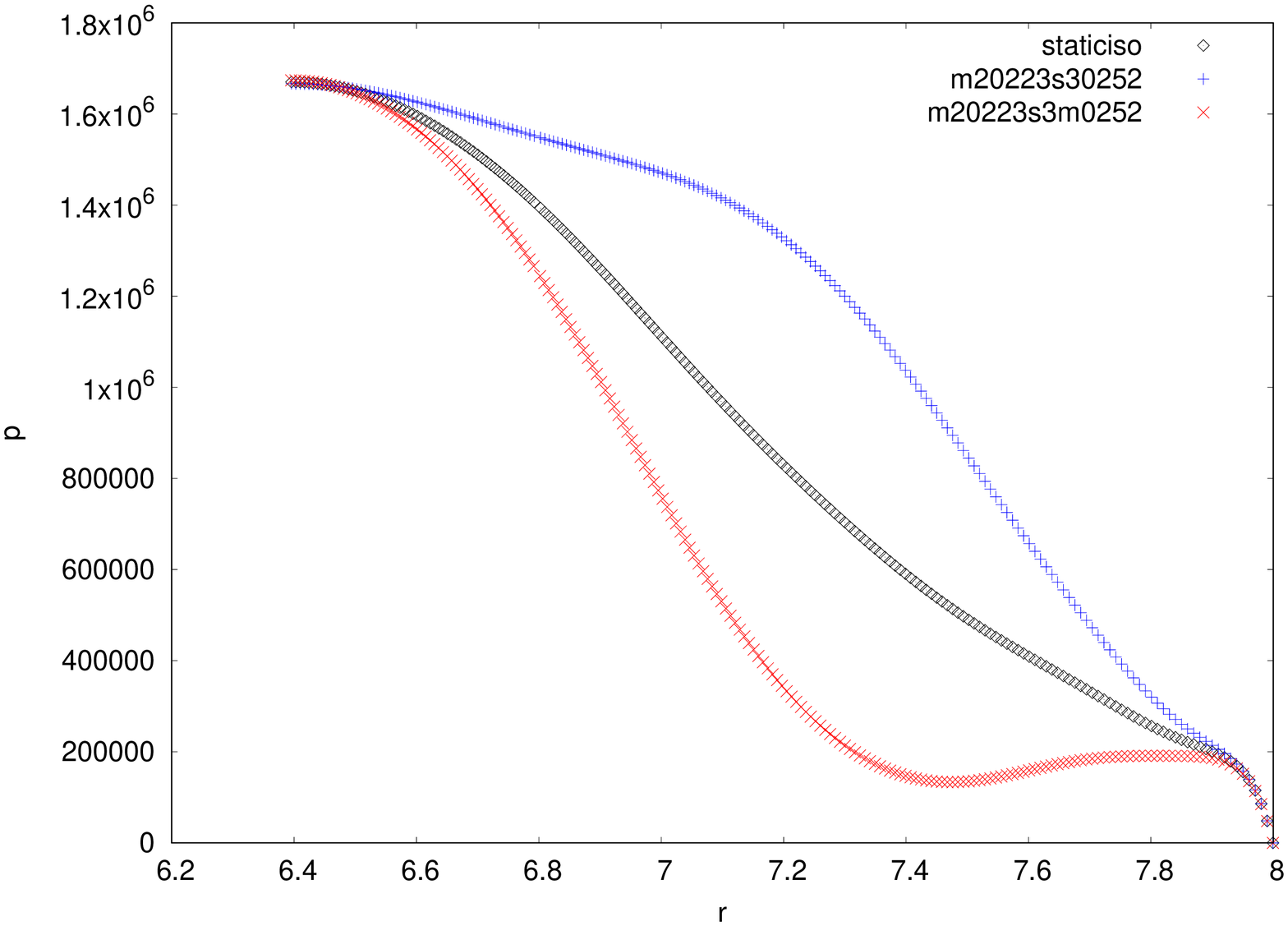}
\caption{Radial profiles of the parallel pressure on the mid-plane $z=0$ for various cases of the pressure anisotropy 
and plasma rotation. Case 2: (blue \textcolor{blue}{$+$}) peaked on-axis rotation with $M_0=0.02$, $m=2$, $n=3$ and 
peaked off-axis pressure anisotropy with $\sigma_0=0.02$, $k=5$, $\ell=2$; Case 3: (red \textcolor{red}{$\times$}) 
peaked on-axis rotation with $M_0=0.02$, $m=2$, $n=3$ and peaked off-axis pressure anisotropy with $\sigma_0=-0.02$, 
$k=5$, $\ell=2$. The static effective pressure (black $\diamondsuit$, Case 1) is plotted for reference.}
\label{fig:pres3}
\end{center}
\end{figure}

\begin{figure}[ht!]
\begin{center}
\psfrag{p}{$p_\perp$ (Pa)}
\psfrag{r}{R(m)}
\psfrag{staticiso }{\small{Case 1}}
\psfrag{m30452s20223 }{\hspace{0.8cm}\small{Case 2}}
\psfrag{m30452s2m0223 }{\hspace{1cm}\small{Case 3}}
\includegraphics[scale=0.46]{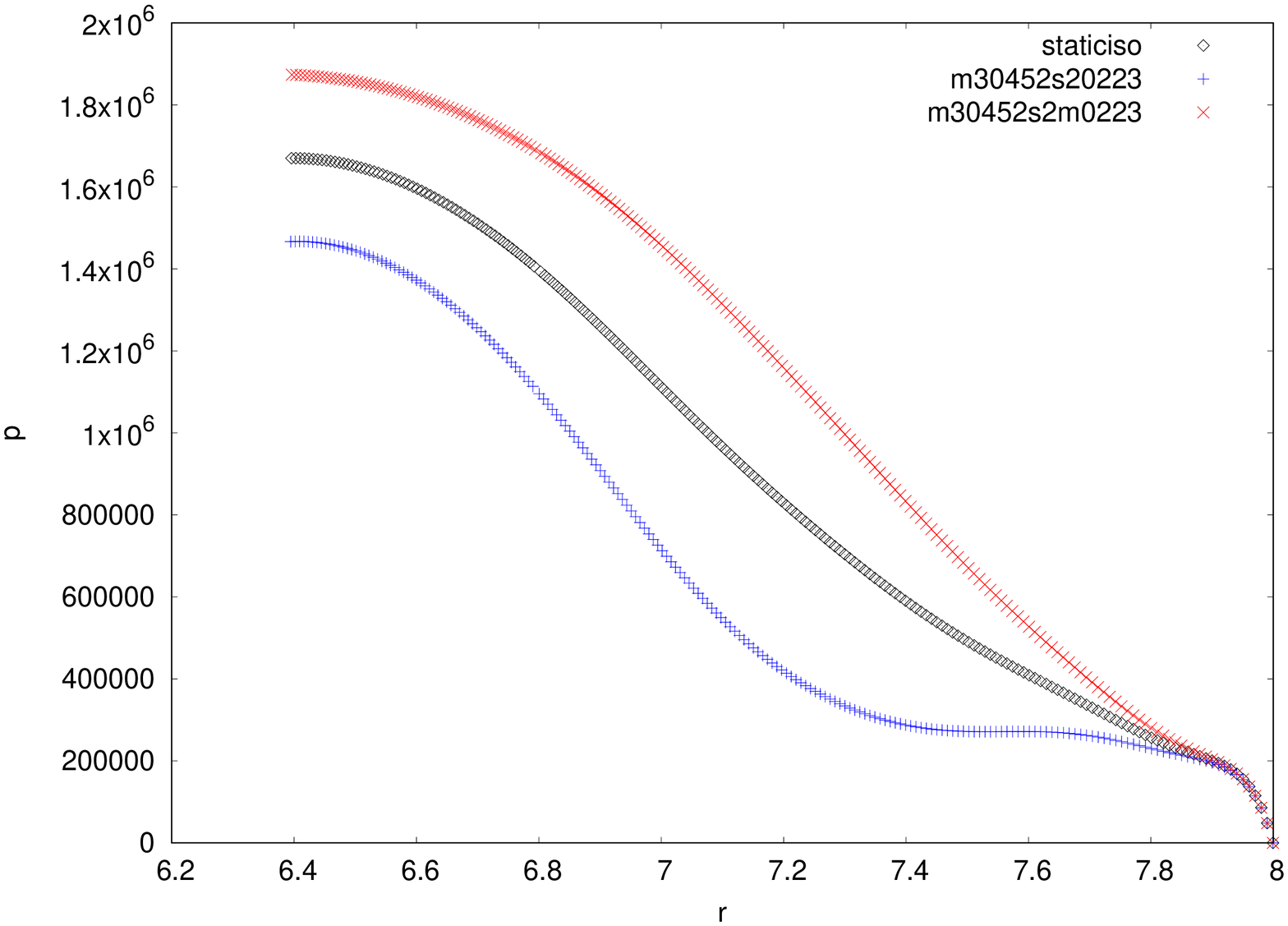}
\caption{Radial profiles of the perpendicular pressure on the mid-plane $z=0$ for various cases of the pressure anisotropy 
and plasma rotation. Case 2: (blue \textcolor{blue}{$+$}) peaked off-axis rotation with $M_0=0.04$, $m=5$, $n=2$ and 
peaked on-axis pressure anisotropy with $\sigma_0=0.02$, $k=2$, $\ell=3$; Case 3: (red \textcolor{red}{$\times$}) 
peaked off-axis rotation with $M_0=0.04$, $m=5$, $n=2$ and peaked on-axis pressure anisotropy with $\sigma_0=-0.02$, 
$k=2$, $\ell=3$. The static effective pressure (black $\diamondsuit$, Case 1) is plotted for reference.}
\label{fig:pres6}
\end{center}
\end{figure}

\begin{figure}[ht!]
\begin{center}
\psfrag{p}{$p_\perp$ (Pa)}
\psfrag{r}{R(m)}
\psfrag{staticiso }{\small{Case 1}}
\psfrag{m20223s20223 }{\hspace{0.8cm}\small{Case 2}}
\psfrag{m20223s2m0223 }{\hspace{1cm}\small{Case 3}}
\includegraphics[scale=0.46]{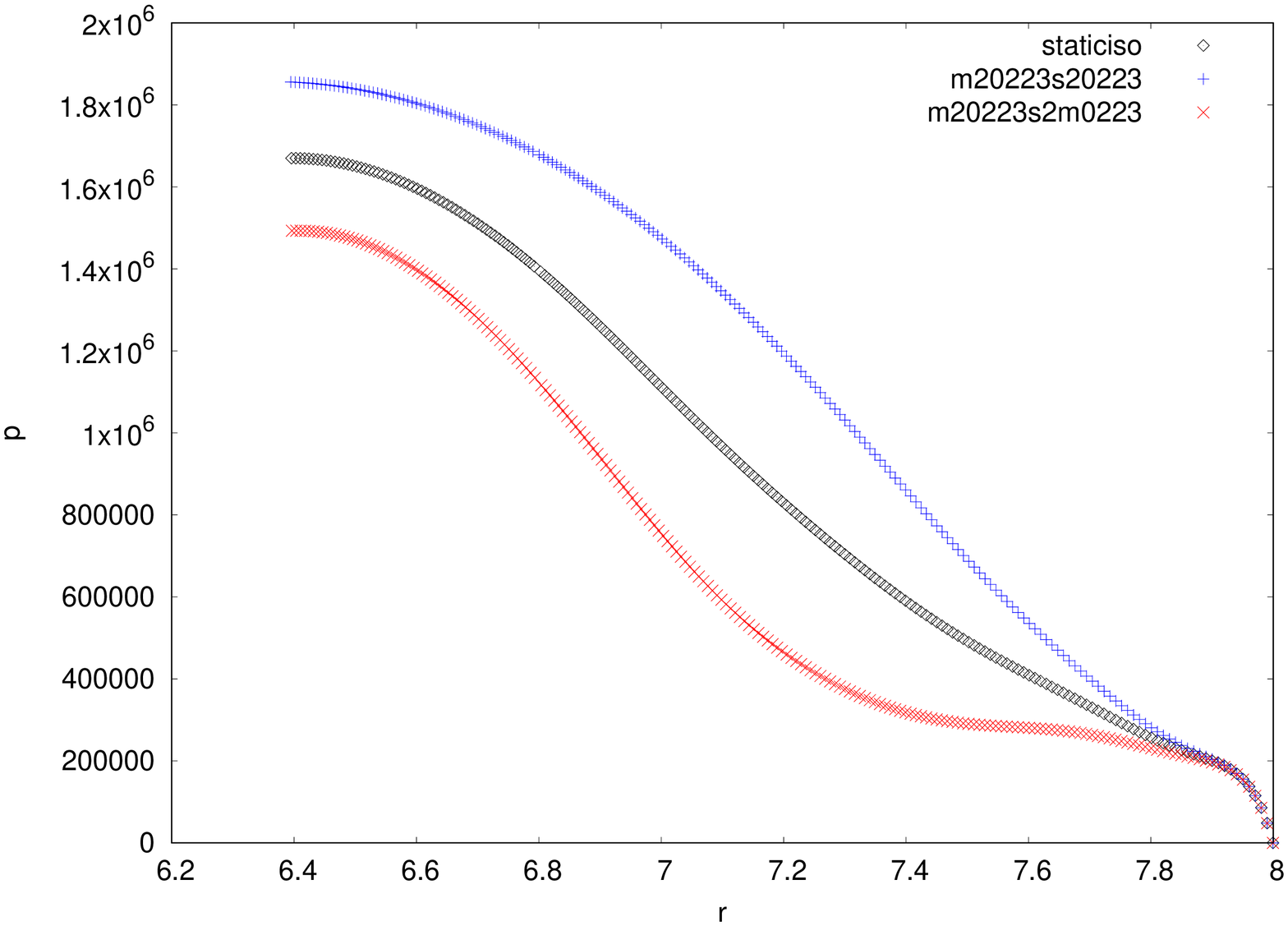}
\caption{Radial profiles of the perpendicular pressure on the mid-plane $z=0$ for various cases of the pressure anisotropy 
and plasma rotation. Case 2: (blue \textcolor{blue}{$+$}) peaked off-axis rotation with $M_0=0.02$, $m=2$, $n=3$ and 
peaked on-axis pressure anisotropy with $\sigma_0=0.02$, $k=2$, $\ell=3$; Case 3: (red \textcolor{red}{$\times$}) 
peaked off-axis rotation with $M_0=0.02$, $m=2$, $n=3$ and peaked on-axis pressure anisotropy with $\sigma_0=-0.02$, 
$k=2$, $\ell=3$. The static effective pressure (black $\diamondsuit$, Case 1) is plotted for reference.}
\label{fig:pres9}
\end{center}
\end{figure}

\begin{figure}[ht!]
\begin{center}
\psfrag{p}{$p_\perp$ (Pa)}
\psfrag{r}{R(m)}
\psfrag{staticiso }{\small{Case 1}}
\psfrag{m30424s30252 }{\hspace{0.8cm}\small{Case 2}}
\psfrag{m30424s30241 }{\hspace{0.8cm}\small{Case 3}}
\includegraphics[scale=0.46]{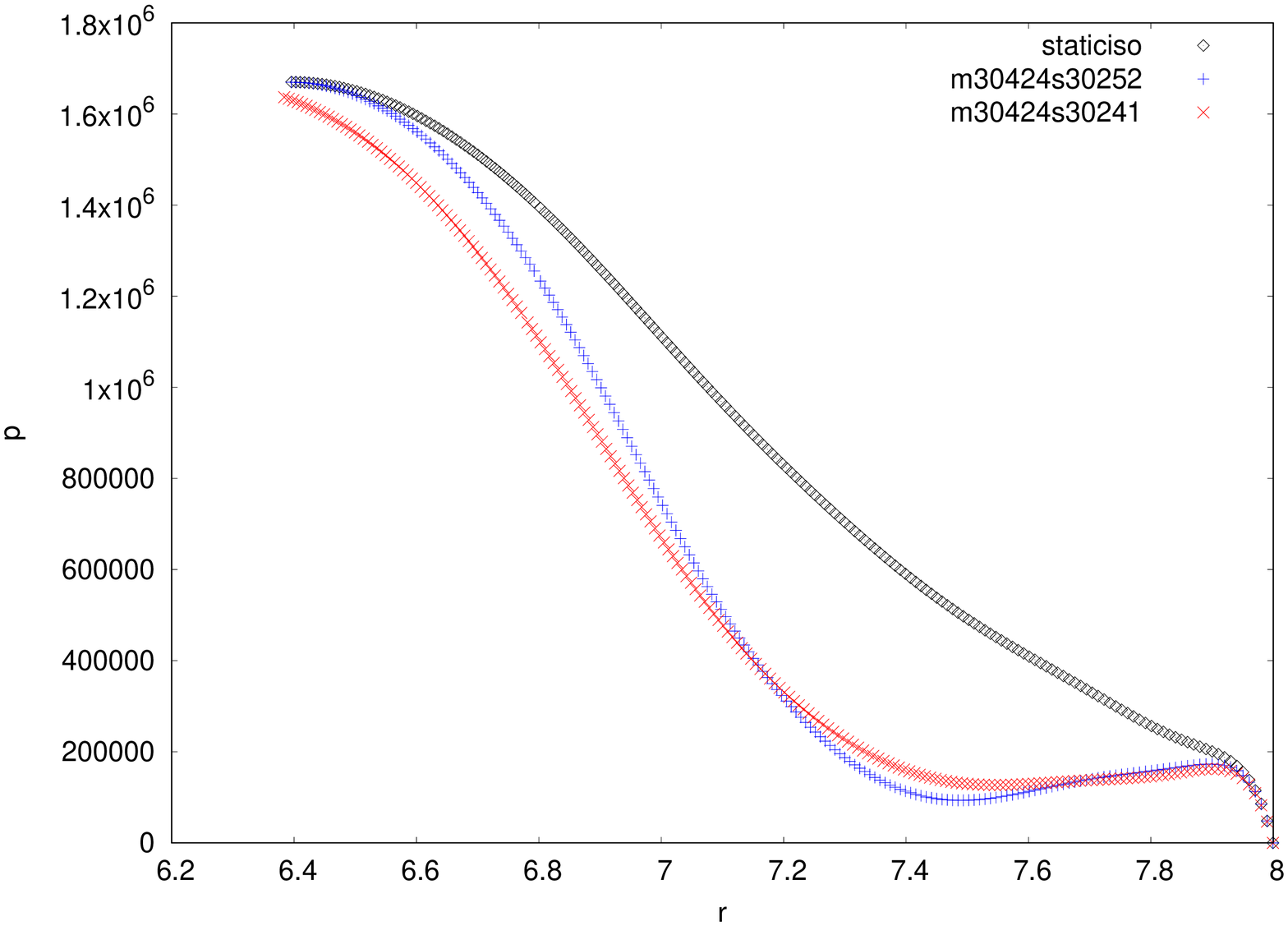}
\caption{Graphs of the perpendicular pressure components profile in the radial direction at $z=0$ for various cases of the pressure anisotropy 
and plasma rotation. Case 2: (blue \textcolor{blue}{$+$}) peaked off-axis plasma rotation with $M_0=0.04$, $m=2$, $n=4$ and peaked off-axis 
pressure anisotropy with $\sigma_0=0.02$, $k=5$, $\ell=2$; Case 3: (red \textcolor{red}{$\times$}) rotation as in Case 2 and peaked off-axis 
pressure anisotropy with $\sigma_0=0.02$, $k=4$, $\ell=1$. The static effective pressure (black $\diamondsuit$, Case 1) is plotted for reference.}
\label{fig:pres7}
\end{center}
\end{figure}

By inspection of Eq. (\ref{eq:11}) is expected that for given $p_s$, plasma rotation and pressure anisotropy in the 
case of $\sigma > 0$ reduce the effective pressure values compared to the static and isotropic one (Fig. \ref{fig:pres1}).
The fact that now there are two independently varied quantities ($M$, $\sigma$) and one of them can be negative, permits 
us to shape the profile with great flexibility. It is interesting that the shear of $\bar{p}$ depends more on the maximum value of the pressure 
anisotropy and less on the values of $k$ and $\ell$. It must be noted here that in general the impact of the pressure anisotropy on the effective 
 pressure is negligible. Nevertheless, we present the observed results for completion. As is evident from Figs. \ref{fig:pres1} and \ref{fig:pres8} 
 the peaked off-axis profile of the pressure anisotropy affects in greater extent the effective pressure profile compared to the peaked on-axis one.
Next, we will focus on the pressure components. It appears that for peaked on-axis $\sigma$, the parallel pressure profile 
flattens on the core, where the maximum value of $\sigma$ is located, thus helping in reducing pressure-gradient-driven modes 
while the higher the values of $\sigma_0$, $k$ and $\ell$, the larger the shear of the pressure component, in a region within 
the plasma volume, as is shown in  Fig. \ref{fig:pres2}. In addition, the impact of plasma rotation has a weak effect on the 
components of the pressure as is evident in the same Fig. \ref{fig:pres2}. Pressure profiles with high shear region in the midplane 
of the poloidal cross-section are observed in discharges with ITBs. 
As already mentioned above the fact that in the presence of anisotropy, $\sigma$ can be either positive or negative 
(see \cite{2001PPCF...43.1441Z}, \cite{2011PPCF...53g4021H}), while it vanishes for isotropic pressure, allows one to 
explore additional possible experimental configurations, in connection to the direction of the applied heating. The 
sign of $\sigma$ appears to affect equally the parallel and perpendicular components of the pressure as it is evident 
in Figs. \ref{fig:pres3}, \ref{fig:pres6} and \ref{fig:pres9}, regardless of whether the maximum of $\sigma$ is located 
at the magnetic axis or at another point within the plasma volume, thus not favouring a specific set-up. 
In general, the effect of pressure anisotropy on the pressure components is qualitatively the same, though reversed. As is evident 
from Figures \ref{fig:pres3} and \ref{fig:pres9} for $\sigma>0$ $p_\parallel$ increases while $p_\perp$ reduces, 
compared to the isotropic case. The impact of pressure anisotropy appears to be stronger for peaked off-axis $\sigma$-profiles compared 
to cases with peaked on-axis profiles (Figs. \ref{fig:pres3}, \ref{fig:pres9}), a result similar to that for isotropic plasmas 
with parallel rotation \citep{2016PhPl...23g2507P}. In the case of isotropic plasmas with parallel rotation the experimental 
values of $M$ in order to obtain  pressure profiles similar to those observed in discharges with ITBs or H-mode plasmas, i.e. 
profiles with steep regions in the vicinity  of the barrier and associated with the maximum of the Mach number profile, appear to be 
difficult to achieve, especially in large devices. On the contrary, pressure anisotropy which gives similar results for the pressure 
components profiles is achievable. For peaked  off-axis $\sigma$-profiles the impact to the pressure components profiles is stronger 
in the case where the position of $\sigma_0$ is closer to the magnetic axis (Fig. \ref{fig:pres7}).
\begin{figure}[ht!]
\begin{center}
\psfrag{p}{$p_\parallel$ (Pa)}
\psfrag{r}{R(m)}
\psfrag{staticiso }{\small{Case 1}}
\psfrag{m0s20223 }{\hspace{0.25cm}\small{Case 2}}
\psfrag{m20223s20223 }{\hspace{0.8cm}\small{Case 3}}
\psfrag{m20223s20226 }{\hspace{0.8cm}\small{Case 4}}
\includegraphics[scale=0.46]{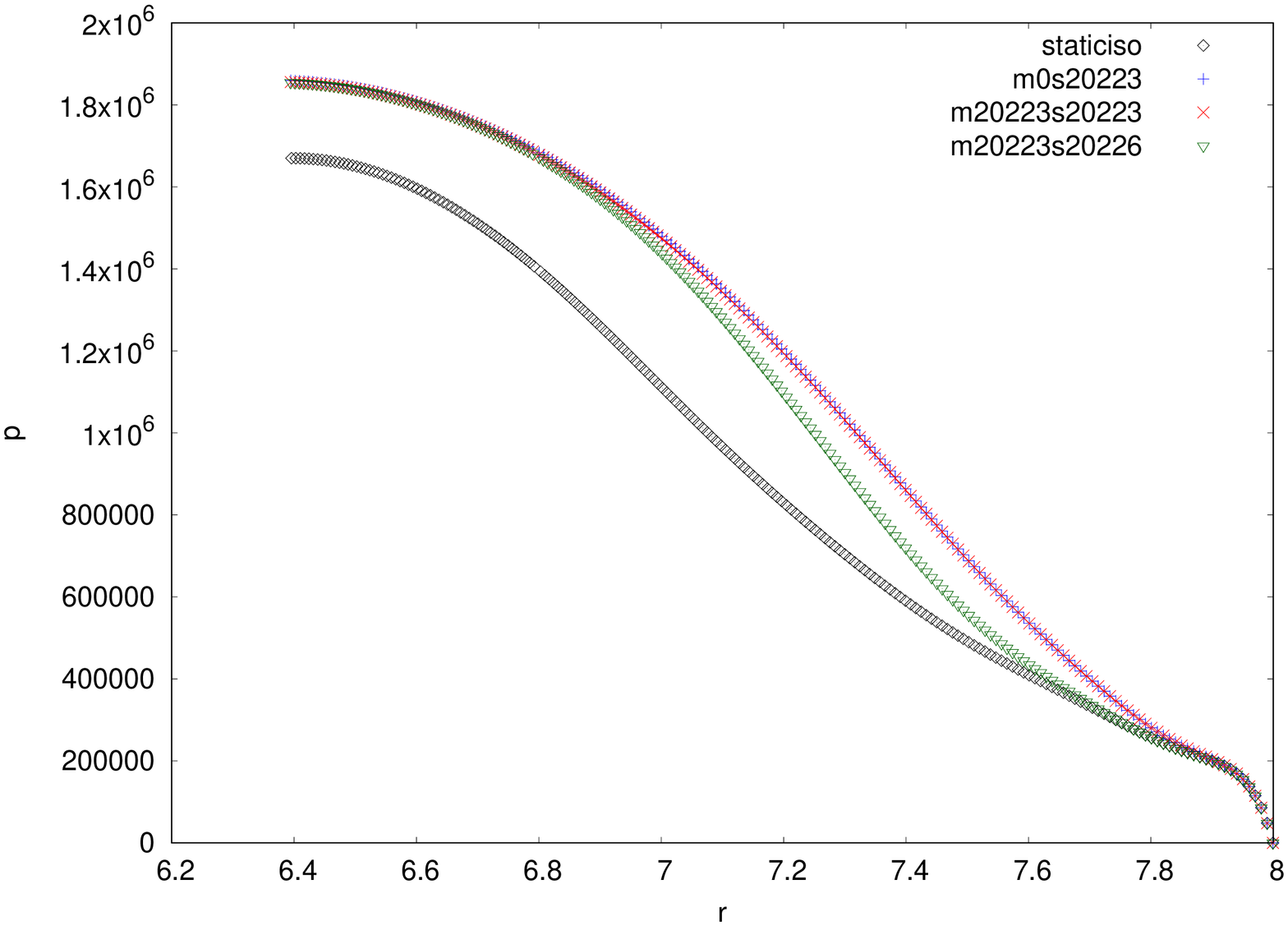}
\caption{Graphs of the parallel pressure components profile in the radial direction at $z=0$ for various cases of the pressure 
anisotropy and plasma rotation. Case 2: (blue \textcolor{blue}{$+$}) no rotation and peaked on-axis pressure anisotropy with 
$\sigma_0=0.02$, $k=2$, $\ell=3$; Case 3: (red \textcolor{red}{$\times$}) peaked on-axis rotation with $M_0=0.02$, $m=2$, $n=3$ 
and peaked on-axis pressure anisotropy with $\sigma_0=0.02$, $k=2$, $\ell=3$; Case 4: (green \textcolor{green}{$\triangledown$}) 
peaked on-axis rotation with $M_0=0.02$, $m=2$, $n=3$ and peaked on-axis pressure anisotropy with $\sigma_0=0.02$, $k=2$, $\ell=6$. 
The static effective pressure (black $\diamondsuit$, Case 1) is plotted for reference.}
\label{fig:pres4}
\end{center}
\end{figure}

\begin{figure}[ht!]
\begin{center}
\psfrag{j}{$J_\parallel$ (A/$m^2$)}
\psfrag{r}{R(m)}
\psfrag{staticiso }{\small{Case 1}}
\psfrag{m30452s20223 }{\hspace{0.8cm}\small{Case 2}}
\psfrag{m30452s2m0223 }{\hspace{1cm}\small{Case 3}}
\psfrag{m0s20223 }{\hspace{0.25cm}\small{Case 4}}
\includegraphics[scale=0.46]{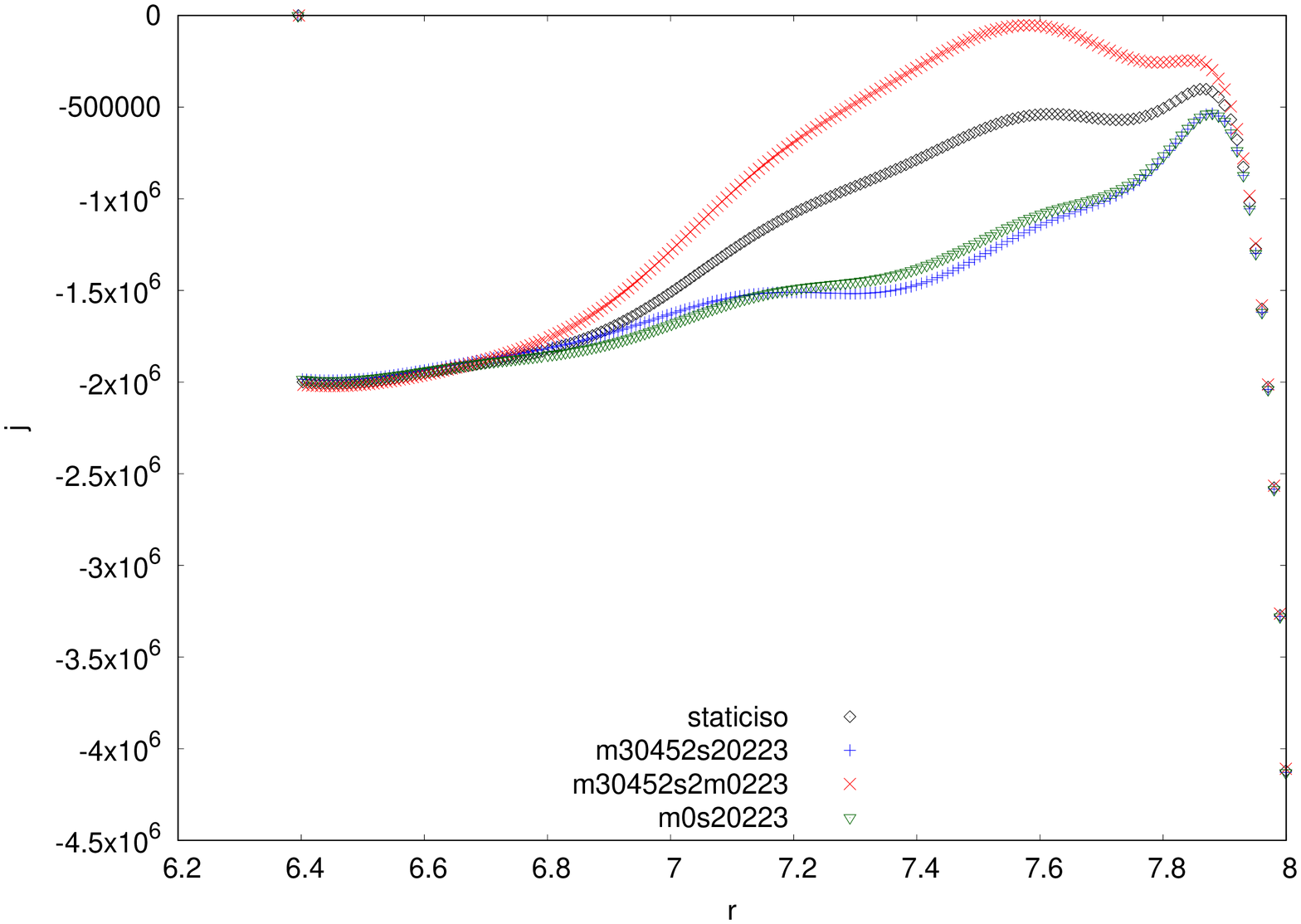}
\caption{Plots of the parallel to the magnetic field current density versus the radial distance from 
the axis of symmetry on the mid-plane $z=0$ for cases of pressure anisotropy peaked on-axis and parallel rotation peaked off-axis.  
Case 2: (blue \textcolor{blue}{$+$}) peaked off-axis rotation with $M_0=0.04$, $m=5$, $n=2$ and peaked on-axis pressure anisotropy 
with $\sigma_0=0.02$, $k=2$, $\ell=3$; Case 3: (red \textcolor{red}{$\times$}) peaked off-axis rotation with $M_0=0.04$, $m=5$, $n=2$ 
and peaked on-axis pressure anisotropy with $\sigma_0=-0.02$, $k=2$, $\ell=3$; Case 4: (green \textcolor{green}{$\triangledown$}) 
no rotation and peaked on-axis pressure anisotropy with $\sigma_0=0.02$, $k=2$, $\ell=3$. The case of static and isotropic 
(black $\diamondsuit$, Case 1) is plotted for reference.}
\label{fig:jpar1}
\end{center}
\end{figure}

\begin{figure}[ht!]
\begin{center}
\psfrag{j}{$J_\parallel$ (A/$m^2$)}
\psfrag{r}{R(m)}
\psfrag{staticiso }{\small{Case 1}}
\psfrag{m20223s30252 }{\hspace{0.8cm}\small{Case 2}}
\psfrag{m20223s3m0252 }{\hspace{1cm}\small{Case 3}}
\psfrag{m0s30252 }{\hspace{0.25cm}\small{Case 4}}
\includegraphics[scale=0.46]{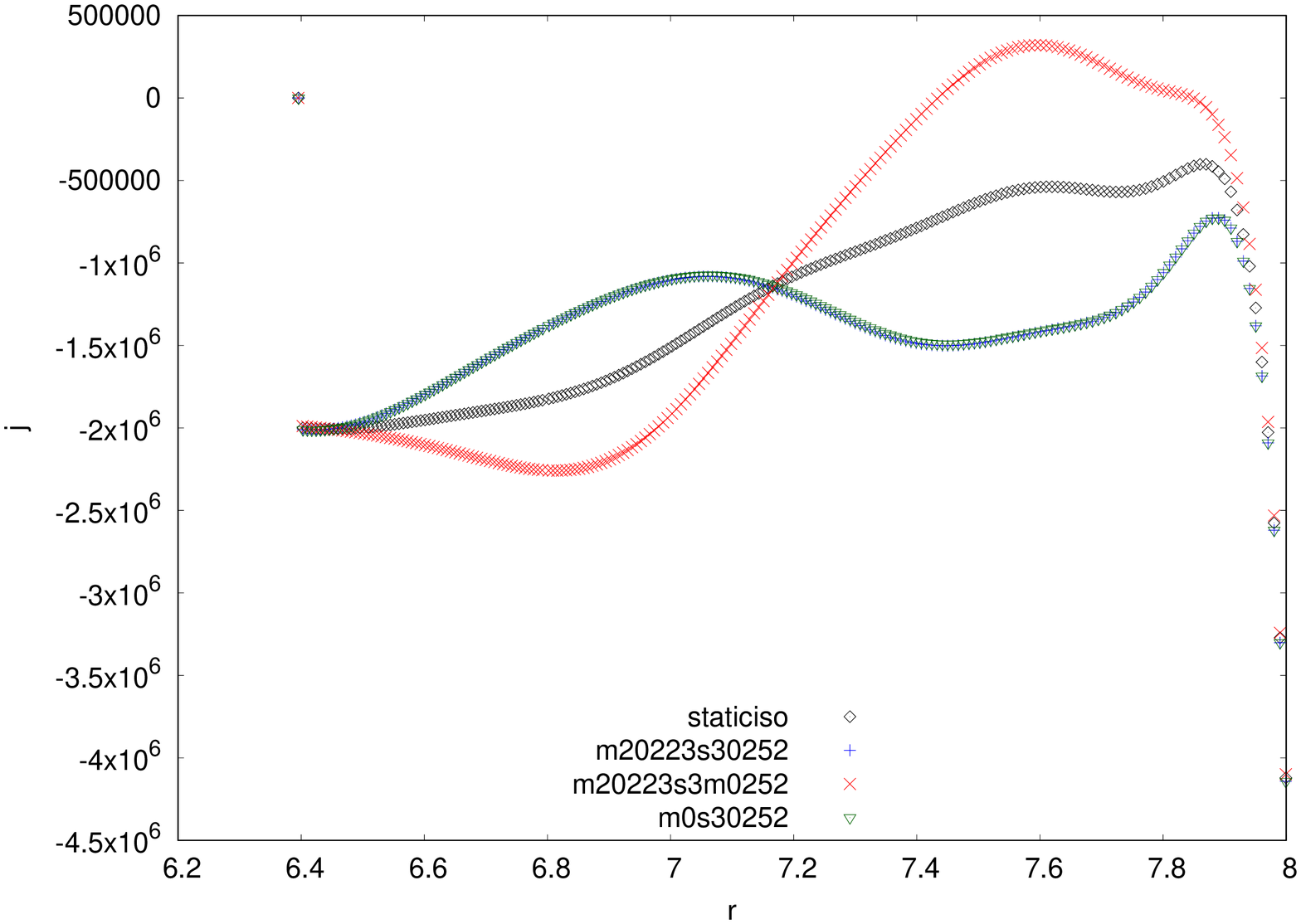}
\caption{Plots of the parallel to the magnetic field current density versus the radial distance from 
the axis of symmetry on the mid-plane $z=0$ for cases of pressure anisotropy peaked off-axis and parallel rotation peaked on-axis. 
Case 2: (blue \textcolor{blue}{$+$}) peaked on-axis rotation with $M_0=0.02$, $m=2$, $n=3$ and peaked off-axis pressure anisotropy 
with $\sigma_0=0.02$, $k=5$, $\ell=2$; Case 3: (red \textcolor{red}{$\times$}) peaked on-axis rotation with $M_0=0.02$, $m=2$, $n=3$ 
and peaked off-axis pressure anisotropy with $\sigma_0=-0.02$, $k=5$, $\ell=2$; Case 4: (green \textcolor{green}{$\triangledown$}) 
no rotation and peaked off-axis pressure anisotropy with $\sigma_0=0.02$, $k=5$, $\ell=2$.  The case of static and isotropic (black 
$\diamondsuit$, Case 1) is plotted for reference.}
\label{fig:jpar2}
\end{center}
\end{figure}

\begin{figure}[ht!]
\begin{center}
\psfrag{j}{$J_\parallel$ (A/$m^2$)}
\psfrag{r}{R(m)}
\psfrag{staticiso }{\small{Case 1}}
\psfrag{m30424s30252 }{\hspace{0.8cm}\small{Case 2}}
\psfrag{m30424s3m0252 }{\hspace{1cm}\small{Case 3}}
\psfrag{m30424s30241 }{\hspace{0.8cm}\small{Case 4}}
\includegraphics[scale=0.46]{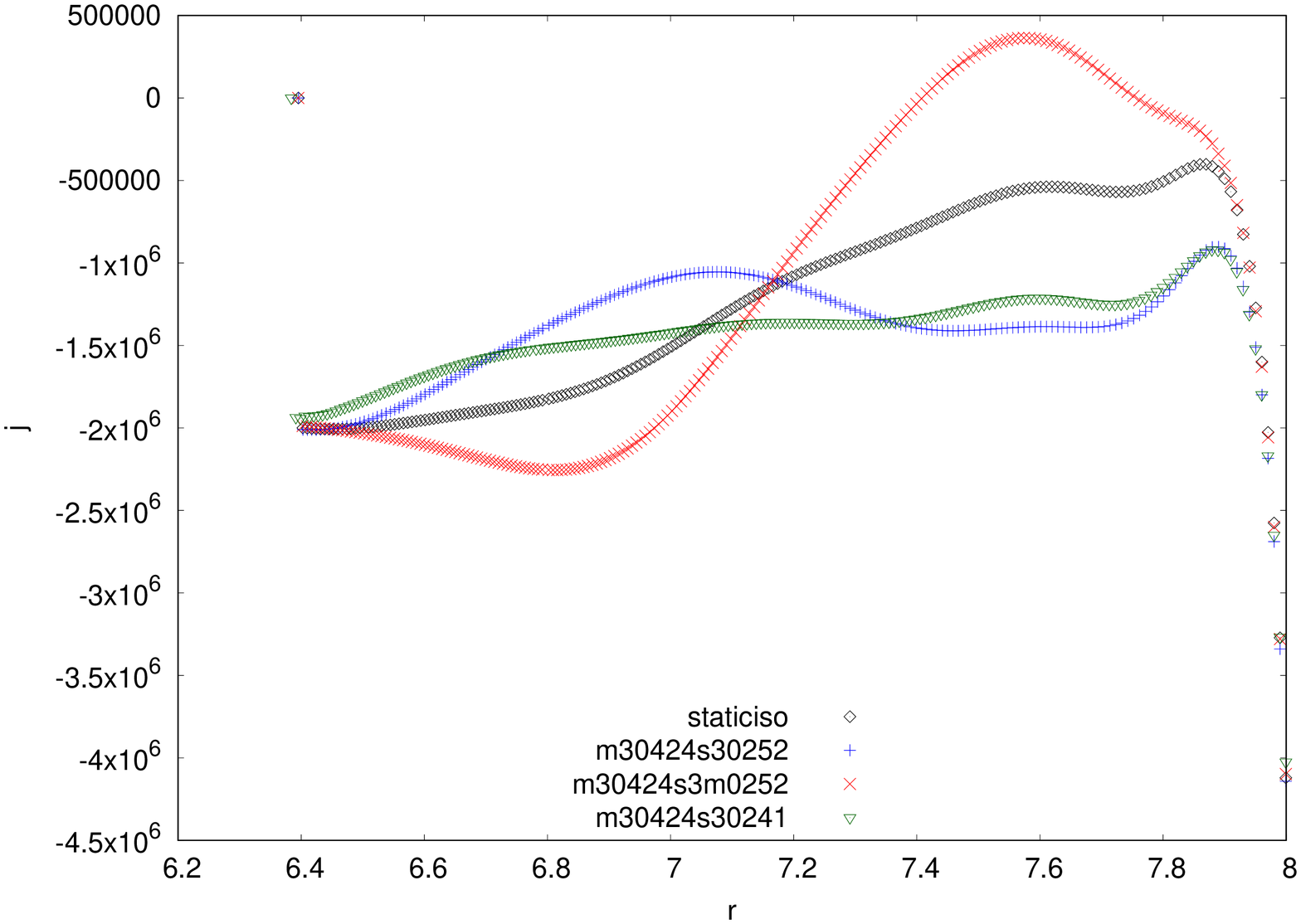}
\caption{Plots of the parallel to the magnetic field current density versus the radial distance from 
the axis of symmetry on the mid-plane $z=0$ for cases of pressure anisotropy peaked on-axis and parallel rotation peaked off-axis.  
Case 2: (blue \textcolor{blue}{$+$}) peaked off-axis rotation with $M_0=0.04$, $m=2$, $n=4$ and peaked off-axis pressure anisotropy 
with $\sigma_0=0.02$, $k=5$, $\ell=2$; Case 3: (red \textcolor{red}{$\times$}) peaked off-axis rotation with $M_0=0.04$, $m=2$, 
$n=4$ and peaked off-axis pressure anisotropy with $\sigma_0=-0.02$, $k=5$, $\ell=2$; Case 4: (green \textcolor{green}{$\triangledown$}) 
peaked off-axis rotation with $M_0=0.04$, $m=2$, $n=4$ and peaked off-axis pressure anisotropy with $\sigma_0=0.02$, $k=4$, 
$\ell=1$. The case of static and isotropic (black $\diamondsuit$, Case 1) is plotted for reference.}
\label{fig:jpar3}
\end{center}
\end{figure}

\begin{figure}[ht!]
\begin{center}
\psfrag{j}{$J_\parallel$ (A/$m^2$)}
\psfrag{r}{R(m)}
\psfrag{staticiso }{\small{Case 1}}
\psfrag{m20223s20223 }{\hspace{0.8cm}\small{Case 2}}
\psfrag{m20223s20226 }{\hspace{0.8cm}\small{Case 3}}
\psfrag{m0s2m0223 }{\hspace{0.45cm}\small{Case 4}}
\includegraphics[scale=0.46]{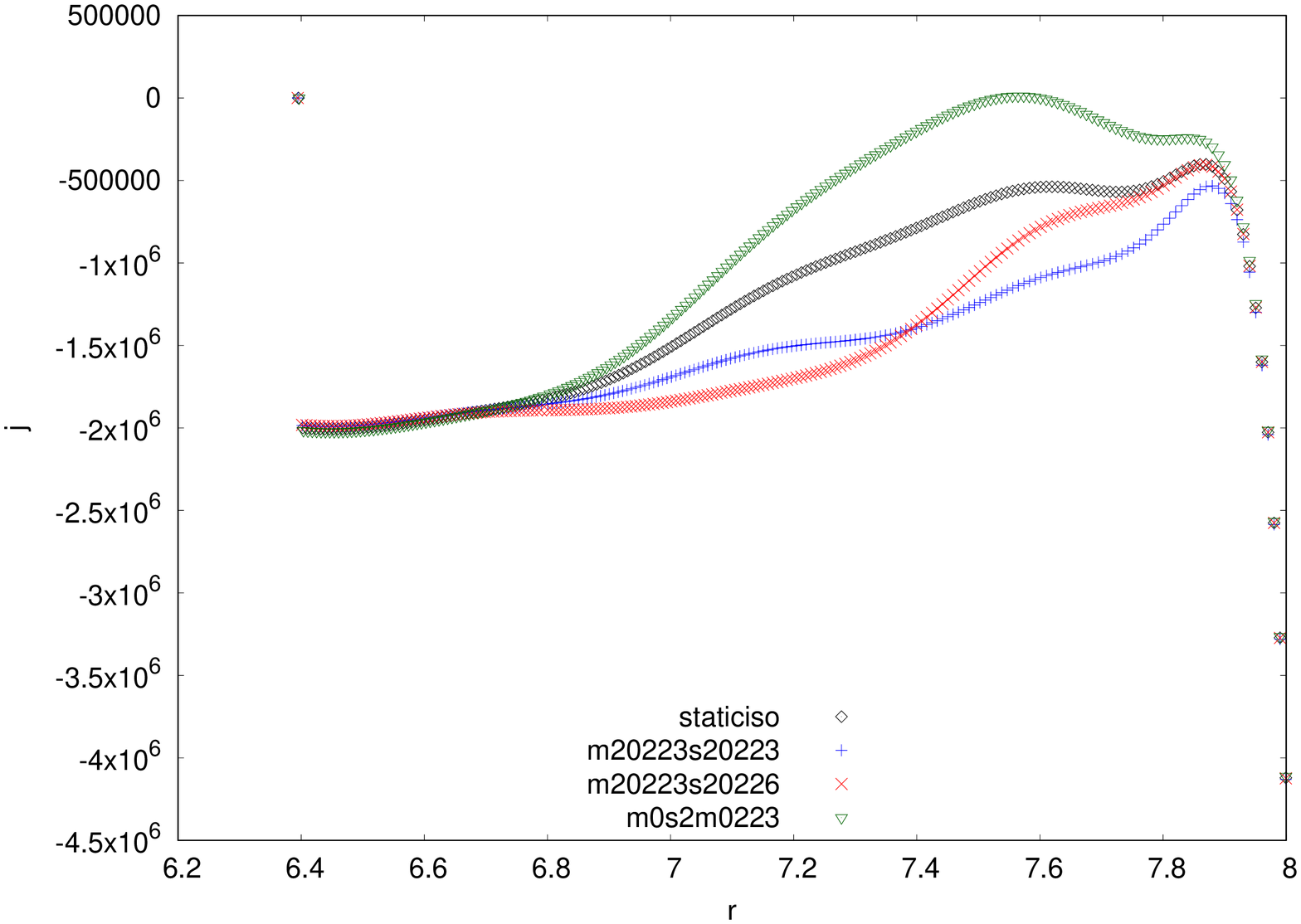}
\caption{Plots of the parallel to the magnetic field current density versus the radial distance from the axis of symmetry 
on the mid-plane $z=0$ for cases of pressure anisotropy peaked off-axis and parallel rotation peaked on-axis. Case 2: 
(blue \textcolor{blue}{$+$}) peaked on-axis rotation with $M_0=0.02$, $m=2$, $n=3$ and peaked on-axis pressure anisotropy 
with $\sigma_0=0.02$, $k=2$, $\ell=3$; Case 3: (red \textcolor{red}{$\times$}) peaked on-axis rotation with $M_0=0.02$, $m=2$, $n=3$ 
and peaked on-axis pressure anisotropy with $\sigma_0=0.02$, $k=2$, $\ell=6$; Case 4: (green \textcolor{green}{$\triangledown$}) 
no rotation and peaked on-axis pressure anisotropy with $\sigma_0=-0.02$, $k=2$, $\ell=3$. The case of static and isotropic 
(black $\diamondsuit$, Case 1) is plotted for reference.}
\label{fig:jpar4}
\end{center}
\end{figure}

Paying attention to the current density, we must note that for the specific input, the impact on both the parallel and toroidal current 
density profiles is the same, and therefore we present as examples only of one of the components. The overall conclusion is that 
the pressure anisotropy can affect the current density much stronger than the parallel rotation. In addition, the impact is 
localized in the region where the shear of $\sigma$ is located. Therefore, for peaked on-axis $\sigma$-profiles (Fig. 
\ref{fig:jpar1}), the impact on the current density is located towards the edge of the poloidal cross-section, with the extent 
of this region being dependent on the shape of the $\sigma$ profile. For peaked off-axis $\sigma$-profiles, the current density 
is affected almost throughout the poloidal cross-section except for the $\sigma_0$ point, the boundary and the magnetic axis 
(Fig. \ref{fig:jpar2}). The region of strongest impact is in the middle of the distance between the point of maximum $\sigma$ 
and the magnetic axis or the boundary. The above results are similar with the ones obtained in \citep{2016PhPl...23g2507P} for the 
impact of parallel rotation on the current density. Compared to the case of parallel rotation, the impact of pressure anisotropy on 
the current density is apparently more profound, (Figs. \ref{fig:jpar1} and \ref{fig:jpar2}). 
As can be seen from Figs. \ref{fig:jpar3} and \ref{fig:jpar4} for $\sigma<0$, the current density profiles have increased 
gradient therefore affecting the relevant modes. Consequently, from the stability point-of-view, it appears that heating parallel to
the magnetic surfaces is desirable since it will smooth out the current density profiles. Moreover, the smaller the shear of 
$\sigma$, the smaller the current density gradient. 
Additionally, regardless of whether it is favourable or not from the stability point-of-view, for peaked 
on-axis $\sigma$-profiles the effect on the current density is stronger for $\sigma<0$ (Fig. \ref{fig:jpar2}) while the 
opposite is observed for peaked off-axis $\sigma$-profiles (Fig. \ref{fig:jpar4}). 

Regarding the safety factor, though the transformation does not affect its values (cf. Eqs. (\ref{eq:q1}) and (\ref{eq:q2})), the 
rotation and the pressure anisotropy alter the $q$-values through their impact on the input data. Specifically, for $\sigma>0$, 
$q$ increases in the region where the anisotropy is localized. For example, in the case of static anisotropic equilibrium 
with $\sigma$ peaked on-axis the values of $q$ at the magnetic axis get larger, as was also reported in \citep{2011PPCF...53g4021H} 
while at the boundary either get larger or smaller. For $\sigma<0$ and peaked on-axis, $q$ decreases thereon. The picture is 
different for $\sigma$-profiles peaked off-axis, i.e. for all the cases examined, the values of $q$ decrease at the magnetic 
axis as well as at the boundary. As in the case of peaked on-axis $\sigma$-profiles, just the opposite occurs for $\sigma<0$. 

Examining the impact of the pressure anisotropy on the position of the magnetic axis, we found an inward shift for peaked on-axis 
profiles and $\sigma>0$, a similar result found also in \citep{2001PPCF...43.1441Z}, and an outward shift for $\sigma<0$. For 
peaked off-axis $\sigma$ the position remains practically the same. These results regarding the safety factor and the position 
of the magnetic axis were not observed in \citep{2016PhPl...23g2507P} due to the weak impact of parallel rotation on these 
quantities, especially compared to the respective impact of pressure anisotropy. Finally, we examined the impact of pressure 
anisotropy on the toroidal $\beta$, concluding that for positive $\sigma$ a slight decrease is observed in its values for 
both peaked on-axis and peaked off-axis profiles for the anisotropy.  Note that the toroidal $\beta$ in the table \ref{tab:1} 
is defined as
\begin{equation}
 \beta_t=\frac{<p>}{B_0^2/2\mu_0},
 \label{eq:beta}
\end{equation}
where
$$
<p>=\frac{\int_0^{V_0} pdV}{V_0},
$$
$B_0$ the vacuum magnetic field at the geometrical center and $V_0$ the 
total plasma volume.
\begin{table}
\centering
\caption{Values of the safety factor and the toroidal $\beta$ for various cases of rotation and pressure anisotropy.}
\begin{tabular}{| c | c | c | c | c | c | c | c | c | c |}
\hline \hline
q & $\beta_t$ & $M_0$ & $m$ & $n$ & Type & $\sigma_0$ & $k$ & $\ell$ & Type \\
\hline
 0.6478 & 0.04199 & 0 & - & - & - & 0.02 & 2 & 3 & on-axis \\
\hline
 0.6491 & 0.04199 & 0 & - & - & - & 0.02 & 2 & 6 & on-axis \\
\hline
 0.6517 & 0.04197 & 0 & - & - & - & 0.035 & 2 & 6 & on-axis \\
\hline
 0.6477 & 0.04199 & 0.02 & 2 & 3 & on-axis & 0.02 & 2 & 3 & on-axis \\
\hline
 0.6370 & 0.04204 & 0 & - & - & - & -0.02 & 2 & 3 & on-axis \\
\hline
 0.6387 & 0.04200 & 0 & - & - & - & 0.02 & 5 & 2 & off-axis \\
\hline
 0.6387 & 0.04200 & 0 & - & - & - & 0.02 & 5 & 2 & off-axis \\
\hline
 \end{tabular}
 \label{tab:1}
\end{table}

\section{Conclusions}

We examined the impact of pressure anisotropy and parallel rotation on the equilibrium properties of axisymmetric, toroidally 
confined plasmas by means of numerically constructed ITER-like configurations. To this end the appropriate GGS equation, 
with the aid of an integral transformation is put in a form identical to the well-known GS equation. This transformation maps 
the poloidal flux function $\psi$ to another flux function $u$ preserving the shape of the magnetic surfaces. We modified the 
fixed boundary equilibrium code HELENA, so that via the direct and the inverse transformation the calculated by the Grad-Shafranov 
solver equilibrium quantities, now in the $u$-space, are mapped to the $\psi$-space.

On the basis of the equilibria constructed by the code we examined the impact of pressure anisotropy, rotation and their shear on the 
pressure, toroidal current density, safety factor, position of the magnetic axis and toroidal beta. We mainly focused on the impact of 
the pressure anisotropy, while the rotation was primarily used for comparison since its impact on the equilibrium was examined in a previous 
study \citep{2016PhPl...23g2507P}. The presence of pressure anisotropy in addition to parallel rotation, allows access to configurations 
with higher values of Mach number, for $\sigma<0$, and more capabilities in shaping the equilibrium quantities profiles. The effect 
of pressure anisotropy is much stronger compared to that of parallel rotation on all the quantities we examined, with the exception 
of the effective pressure. As expected, for $\sigma>0$ the effective pressure, which enters the GGS equation, is reduced by the 
pressure anisotropy, though the impact is minimal. The impact on the pressure components of peaked off-axis pressure anisotropy is 
stronger than that for peaked on-axis anisotropy, and especially for pressure anisotropy localized close to the magnetic axis. 
In addition, the impact of pressure anisotropy is the same regardless of the sign of $\sigma$ and also the same for the two pressure 
components. For the current density the peaked off-axis anisotropy has a stronger impact throughout the poloidal cross-section as 
opposed to the peaked on-axis case where the impact is localized close to the magnetic axis. This is due to the fact that the shear 
of pressure anisotropy has a stronger effect on the current density that the anisotropy itself. In this case that impact is direction 
independent, unlike the case of parallel rotation which affects more drastically the parallel current component than the toroidal 
one. The peaked off-axis $\sigma$-profile with low shear is smoothing out the current density profiles thus being favourable from 
the stability point-of-view, since it can affect current-gradient instabilities. It must be noted that, in general $\sigma>0$ 
is beneficial for the configuration because it does not produce large current density gradients. For peaked on-axis $\sigma$, the 
safety factor increases at the magnetic axis, while the effect for peaked off-axis profiles is negligible. Also, in certain cases 
we found that the magnetic axis is shifted inwards for $\sigma>0$  and outward for $\sigma<0$. Finally, the toroidal $\beta$ decreases 
for $\sigma>0$ and  increases for $\sigma<0$. 

As a next step, it is planned to extend further the computation for non-parallel incompressible plasma rotation. In this case 
the rotation is associated with electric fields which are believed to play a role in the transitions to improved confinement 
modes. This can be done on the basis of Eq. (\ref{1}) (or Eq. (\ref{4})) by including the additional electric field dependent 
$R^4$-term therein.

\section{Acknowledgements}
This work has been carried out within the framework of the EUROfusion Consortium and has received funding from (a) the Euratom 
research and training programme 2014-2018 and 2019-2020 under grant agreement No 633053 and (b) the National Program for the Controlled 
Thermonuclear Fusion, Hellenic Republic. The views and opinions expressed herein do not necessarily reflect those of the European Commission.

\bibliography{helenaanis}

\begin{thebibliography}{10}
\bibitem{velocity.scaling}
{For a typical thermonuclear plasma and magnetic field of the order of 1 Tesla,
  the plasma velocity is of the order of $10$ − $10^2$ Km/sec which lies well
  within the first elliptic region.}
\end{thebibliography}

\end{document}